\pdfoutput=1
\documentclass[12pt,a4paper]{amsart}
\usepackage[a4paper,inner=2.5cm,outer=2.5cm,top=2.5cm,bottom=2.5cm]{geometry}
\usepackage{amsmath,amssymb,amsthm,enumerate,mathtools,stmaryrd
}
\usepackage{hyperref}
\hypersetup{colorlinks=true,linkcolor=blue,citecolor=teal,filecolor=magenta,urlcolor=cyan}

\usepackage[oldstyle]{libertinus}
\usepackage{libertinust1math}

\usepackage[T1]{fontenc}
\usepackage[utf8]{inputenc}

\usepackage{makecell}
\usepackage{graphicx}

\usepackage{float}

\usepackage{extarrows}

\usepackage{xcolor}

\newcommand{\C}{\mathbb{C}}

\usepackage{url}

\usepackage[backend=bibtex,style=alphabetic,sorting=anyt,isbn=false,url=true,doi=false,maxalphanames=10,minalphanames=4,mincitenames=4,maxcitenames=10,minnames=4,maxnames=10,giveninits=true,maxbibnames=99]{biblatex}
\renewbibmacro{in:}{\ifentrytype{article}{}{\printtext{\bibstring{in}\intitlepunct}}}
\AtEveryBibitem{\iffieldundef{eprint}{}{\clearfield{url}}}
\DefineBibliographyStrings{english}{page = {},pages = {}}
\DeclareFieldFormat{eprint:arxiv}{\footnotesize
	arXiv:\,\href{https://arxiv.org/abs/\thefield{eprint}}{#1}
}
\DeclareFieldFormat{url}{\footnotesize
	\iffieldundef{doi}{url:\,\url{\thefield{url}}}
	{doi:\,\href{https://doi.org/\thefield{doi}}{\nolinkurl{\thefield{doi}}}}
}

\theoremstyle{plain}
\newtheorem{theorem}{Theorem}[section]
\newtheorem{proposition}[theorem]{Proposition}
\newtheorem{corollary}[theorem]{Corollary}
\newtheorem{lemma}[theorem]{Lemma}

\theoremstyle{definition}
\newtheorem{definition}[theorem]{Definition}
\newtheorem{example}[theorem]{Example}

\makeatletter
\@addtoreset{proofpart}{theorem}
\makeatother
\theoremstyle{remark}
\newtheorem{remark}[theorem]{Remark}

\newcommand{\Z}{\mathbb{Z}}

\newcommand{\cS}{\mathcal{S}}
\newcommand{\cA}{\mathcal{A}}

\newcommand{\cP}{\mathcal{P}}

\newcommand{\cW}{\mathcal{W}}

\DeclareMathOperator{\Aut}{Aut}

\newcommand{\res}{\mathop{\rm res}}

\newcommand{\1}{1\hskip-.95ex1}

\newcommand{\coeff}[2]{\mathop{[{#1}^{#2}]}}

\newcommand{\set}[1]{\llbracket {#1} \rrbracket}

\newcommand{\arctanh}{\operatorname{arctanh}}

\newcommand{\dd}{\,\mathrm{d}}
\newcommand{\strict}{\mathrm{strict}}
\newcommand{\odd}{\mathrm{odd}}
\newcommand{\bK}{\mathbf{K}}

\newcommand{\R}{\mathbb{R}}
\newcommand{\cG}{{\mathcal{G}}}
\newcommand{\bG}{{\mathbb{G}}}
\newcommand{\fA}{{\mathfrak{A}}}
\newcommand{\fB}{{\mathfrak{B}}}
\newcommand{\cY}{\mathcal{Y}}
\newcommand{\bW}{{\mathbb{W}}}
\DeclareMathOperator{\Pf}{Pf}
\newcommand{\p}{\partial}

\newcommand{\xy}{x-y}

\DeclareFontFamily{U}{rcjhbltx}{}
\DeclareFontShape{U}{rcjhbltx}{m}{n}{<->rcjhbltx}{}
\DeclareSymbolFont{hebrewletters}{U}{rcjhbltx}{m}{n}

\DeclareMathSymbol{\shin}{\mathord}{hebrewletters}{152}

\newcounter{pstep}

\numberwithin{equation}{section}

\title[Topological recursion for spin enumeration of hypermaps]{Topological recursion for spin enumeration of hypermaps}

\author[A.~Alexandrov]{A.~Alexandrov}
\address{A.~A.: Center for Geometry and Physics, Institute for Basic Science (IBS), Pohang 37673, Korea
}
\email{alex@ibs.re.kr}

\author[B.~Bychkov]{B.~Bychkov}
\address{B.~B.: Hebrew University of Jerusalem, Einstein Institute of Mathematics, Givat Ram, Jerusalem, 9190401, Israel}
\email{bbychkov@hse.ru}

\author[P.~Dunin-Barkowski]{P.~Dunin-Barkowski}
\address{P.~D.-B.: Faculty of Mathematics, HSE University, Usacheva 6, 119048 Moscow, Russia; HSE--Skoltech International Laboratory of Representation Theory and Mathematical Physics, Skoltech, Bolshoy Boulevard 30 bld. 1, 121205 Moscow, Russia; and NRC “Kurchatov Institute” -- ITEP, 117218 Moscow, Russia}
\email{ptdunin@hse.ru}

\author[M.~Kazarian]{M.~Kazarian}
\address{M.~K.: Faculty of Mathematics, HSE University, Usacheva 6, 119048 Moscow, Russia; and I.M. Krichever Scientific School, Skoltech, Bolshoy Boulevard 30 bld. 1, 121205 Moscow, Russia}
\email{kazarian@mccme.ru}

\author[S.~Shadrin]{S.~Shadrin}
\address{S.~S.: Korteweg-de Vries Institute for Mathematics, University of Amsterdam, Postbus 94248, 1090GE Amsterdam, The Netherlands}
\email{S.Shadrin@uva.nl}	

\begin{document}
	
\begin{abstract} We establish topological recursion for the spin enumeration of hypermaps, or, in the dual language, spin dessins d'enfants. This problem has remarkable features distinguishing it from other applications of topological recursion to enumerative combinatorics. First, the spectral curve is elliptic rather than rational, as in most cases studied before. Second, it contains degenerate ramification points, and one has to apply a version of degenerate topological recursion. Finally, the meromorphic correlator differentials have ``moving'' poles on the antidiagonals between pairs of distinct arguments, and one has to adjust the loop equations to this kind of degeneracy as well.

The computations involve concepts of different kinds, such as BKP integrability, Gaussian integrals, the $\xy$ swap, and symplectic duality.
\end{abstract}
	
\maketitle
	
\setcounter{tocdepth}{2}
\tableofcontents

\section{Introduction}

A hypermap is an embedded bipartite graph with black and white vertices. Assume that a given hypermap $\gamma$ is odd, that is, the valencies of all its vertices, both black and white, and the degrees of all its faces are \emph{odd} (the degree of a face is half the number of its boundary edges). In that case, there is a natural definition of a \emph{sign}, or spin, of a hypermap, $w(\gamma)\in\{1,-1\}$, and we can define the quantities

\begin{equation}\label{eq:spinNumbers}
C_{g}(\mu_1,\dots,\mu_n)=\sum_{\gamma\in{\rm Hmap}^{\rm odd}_{g}(\mu)}
\frac{w(\gamma)\;\alpha ^{|V_{\rm white}(\gamma)|}\;\beta^{|V_{\rm black}(\gamma)|}}{|\Aut(\gamma)|},
\end{equation}
where ${{\rm Hmap}^{\rm odd}_{g}(\mu)}$ is the set of connected odd hypermaps on a genus~$g$ surface with $n$ marked faces of degrees $\mu_1,\dots,\mu_n$; $|V_{\rm black}|$ and $|V_{\rm white}|$ are the numbers of black and white vertices, respectively; and $\alpha$, $\beta$ are formal parameters. Recall that the genus of a map can be recovered from Euler's formula
\begin{equation}\label{eq:euler-genus}
2g-2=|\mu|-\ell(\mu)-|V_{\rm black}|-|V_{\rm white}|,
\end{equation}
where $\ell(\mu)=n$ is the number of parts in the integer partition $\mu$ and $|\mu| = \sum_{i=1}^{\ell(\mu)} \mu_i$.

The actual definition of the sign is reviewed in Appendix~\ref{appx:spin-sign}. Independently of this definition, for computational purposes one can use the Gunningham character formula~\cite{Gun}, which expresses the generating function for these numbers in terms of Schur $Q$-functions:
\begin{align}
F(p_1,p_3,p_5,\dots)&=\sum_{g,~\mu~{\rm odd}}
2^{g-1+\ell(\mu)}%
\hbar^{2g-2+\ell(\mu)}C_{g}(\mu)\frac{p_\mu}{|\Aut(\mu)|},\label{eq:F-def}
\\ Z=\exp F&=\sum_{\lambda\ \strict}
(2\hbar)^{|\lambda|}
\frac{Q_\lambda(1^{\alpha /(2\hbar)})Q_\lambda(1^{\beta/(2\hbar)})}
     {Q_\lambda(\delta)\,2^{\ell(\lambda)}}Q_\lambda(p).
\label{eq:Z-def}
\end{align}
Here $1^\alpha$ stands for the specialization $p_k=\alpha$ for all $k$, and $\delta$ is the specialization $p_k=\delta_{k,1}$. This generating function governs spin enumeration of hypermaps, or, equivalently, dessins d’enfants.

The non-spin analogue of these numbers plays a distinctive role in combinatorics and mathematical physics, and was studied, for instance,  in~\cite{KZ}. In particular, topological recursion~\cite{EO} for these non-spin numbers was proved in \emph{loc.~cit.} Later, the topological recursion statement was extended in~\cite{ACEH,BDKS02} to a larger family of weighted Hurwitz numbers governed by the so-called hypergeometric family of KP tau functions~\cite{KMMM,OrlovScherbin}. The spin analogue of the hypergeometric family is also well studied; see, e.g.,~\cite{Orlov,Lee,MMN,Alexandrov-BGW}. The corresponding version of weighted spin Hurwitz numbers
for the family with completed cycles also satisfies topological recursion, and the proof is parallel to the one in the non-spin case~\cite{GKL,AS23}. For comparison, we review it in Appendix~\ref{apx:HyperGeom} extending the results of~\cite{AS23}.  It turns out, however, that, unlike the non-spin case, for the simplest possible weight chosen as in~\eqref{eq:spinNumbers} the function~\eqref{eq:Z-def} is \emph{not} a specialization of the hypergeometric family. For this reason, the problem studied in this paper has no direct non-spin analogue and requires the development of new techniques.

Let us collect the numbers~\eqref{eq:spinNumbers} into generating series in an alternative way, by means of $n$-point differentials. Using the loop insertion operator $\nabla_X=\sum_{k\geq1,\;k\ \odd} kX^{k-1}dX\;\frac{\partial}{\partial p_k}$, they are defined as
\begin{equation}\label{eq:omegaZ}
\begin{aligned}
\omega_n(X_1,\dots,X_n)&=\sum_{g\ge0}\hbar^{2g-2+n}\omega^{(g)}_{n}
\\&=\nabla_{X_1}\dots\nabla_{X_n}\log(Z)\bigm|_{p=0}+
\delta_{n,2}\left(\tfrac{1}{(X_1-X_2)^2}+\tfrac{1}{(X_1+X_2)^2}\right)d X_1 d X_2.
\end{aligned}
\end{equation}

The main result of this paper is a version of topological recursion for these differentials. This assertion implies, in particular, that the differentials $\omega^{(g)}_n$, originally defined as formal expansions, extend to globally meromorphic differentials on a Cartesian power of a certain algebraic curve, called the spectral curve, which is the same for all $(g,n)$. Moreover, there is a procedure for computing these differentials in closed form inductively in $g$ and~$n$. It turns out that the spectral curve in question is elliptic; it is given by the equation
\begin{equation}\label{eq:spectral-curve-intro}
w^2=z\,(\alpha +\beta\,z)\,(\beta+\alpha \,z),
\end{equation}
and $X$ can be identified as a local coordinate on the spectral curve at the point $(z,w)=(0,0)$ given by
\begin{equation}\label{eq:X-in-z-intro}
X=\frac{z\,(1-z^2)}{w}.
\end{equation}
All differentials are odd with respect to the elliptic involution $\iota:(z,w)\mapsto(z,-w)$ so that they can be represented in the form
\begin{equation}\label{eq:omega-W-eta-intro}
\omega^{(g)}_n=W^{(g)}_n(z_1,\dots,z_n)\;\eta_1\dots\eta_n,
\end{equation}
where $W^{(g)}_n$ is rational and $\eta=\frac{d z}{w}$ is the basic holomorphic differential.
The initial data for the recursion is contained in the unstable differentials explicitly given by
\begin{align}
\omega^{(0)}_1&=\frac{z_1}{X_1^2}d X_1,
\label{eq:omega01}
\\
\omega^{(0)}_2&=\left(\frac{1}{\left(\xi_1-\xi _2\right){}^2}+\frac{1}{\left(\xi _1+\xi _2\right){}^2}\right)\,d \xi_1\,d \xi_2 ,
\quad \xi_i=\frac{z_i}{w_i}.
\label{eq:omega02}
\end{align}

The stable differentials are determined by a version of topological recursion, but it is \emph{not} the standard one introduced by Chekhov--Eynard--Orantin (also known as the CEO topological recursion)~\cite{EO}, for two independent reasons. First, in addition to its eight nondegenerate critical points, the function $X$ has a pole of order~$3$ at infinity, and this degenerate ramification point also contributes to the poles of the differentials. Second, the two-point
differential $\omega^{(0)}_2$ has poles away from the diagonal, so it is not a
Bergman kernel. These off-diagonal singularities of $\omega^{(0)}_2$ produce additional poles of the differentials for $n\ge2$ on the antidiagonals $z_iz_j=1$.

Thus, the differentials $\omega^{(g)}_n$ have exactly $2n+7$ possible poles in each argument for fixed generic values of the remaining arguments: eight nondegenerate critical points of $X$, the point at infinity, and $n-1$ pairs of preimages of $z_i=1/z_j$ under the elliptic cover. The recursion proceeds in two steps. In the first step, we compute the principal parts of the poles of $\omega^{(g)}_n$ in one chosen argument (say, the first one or the last one) using the so-called loop equations. In the second step, we glue the resulting principal parts into a globally meromorphic differential. This globally meromorphic differential is determined uniquely up to a global holomorphic summand, which we fix by requiring its periods to vanish. Alternatively, we can use the standard projection operator of topological recursion, supported at all $2n+7$ points of the residue set, with an arbitrary choice of Bergman kernel. The resulting differential automatically has vanishing residues and periods, so every stable differential $\omega^{(g)}_n$ is the total differential of a meromorphic function.

The most intriguing feature is the definition of the loop equations used in the recursion. We prove that there is a unified way to formulate these equations that applies to all three kinds of poles: nondegenerate zeros of $dX$, infinity, and the antidiagonals. These loop equations are formulated in terms of \emph{symplectic duality}~\cite{BDKS04,ABDKS05} with a suitably chosen transformation weight.  In particular, one can show that the loop equations at the nondegenerate ramification points and at infinity are equivalent to the standard CEO linear and quadratic loop equations, cf.~\cite{BEO, BS} and to the loop equations of the Bouchard--Eynard recursion~\cite{BE-1,BE-2}, respectively. However, these facts are not used in the proof of topological recursion for spin enumeration of hypermaps. Moreover, for the moving poles on the antidiagonals, we have no alternative formulation of the loop equations other than that using symplectic duality.

\subsection{Digression on variation of the concepts of topological recursion} \label{sec: various TR}
We call the resulting version of degenerate topological recursion \emph{Symplectic Topological Recursion}, or SymplTR. It appears not to have occurred in the literature before, but the results of this paper prompt its structural development.

There are various versions of degenerate topological recursion, in addition to the original one developed in~\cite{EO}: the Bouchard--Eynard recursion~\cite{BE-1}, log topological recursion~\cite{Hock-quantum-diLog,ABDKS05}, generalized topological recursion~\cite{ABDKS07}, etc., see also a small survey in~\cite{hock2025quantumcurvescontextsymplectic}. It is natural to trace the relations between these versions of topological recursion analyzing the corresponding loop equations, and  {Symplectic Topological Recursion} serves as a natural common generalization of all these versions that provides a framework for their comparison.

In view of the importance of this new concept and since it is not necessary for the presentation of the proof in the present paper, we postpone its study to a forthcoming publication~\cite{ABDKS-SymplTR}.

\subsection{Plan of the paper}

Let us now review the plan of the proof of topological recursion. In Section~\ref{sec:BKP}, we prove that the series~\eqref{eq:Z-def} is a BKP tau function~\cite{DJKM}. Since this function does not belong to the hypergeometric family for which the BKP integrability is well known~\cite{Orlov}, this fact requires an independent proof. BKP integrability implies that all information about the tau function is contained in the so-called Baker--Akhiezer kernel~$\bK(X_1,X_2)$, which is a bi-half-differential in just two arguments~\cite{DJKM,AM,WY}.
In particular, all $\omega$-differentials are expressed in terms of $\bK$ by a version of the determinantal formulas proved in~\cite{WY}. In fact, we also need an extended version of the determinantal formulas for the extended differentials, and we review them in Appendix~\ref{sec:xy-sympl}.

Next, in Section~\ref{sec:K-Gauss}, we compute the kernel $\bK$ in a closed form. We first obtain it as a series from its definition and the Schur $Q$-function expansion~\eqref{eq:Z-def};  then we represent the resulting expression as a formal four-dimensional integral. Such a representation provides a new tool: the standard stationary phase analysis leads immediately to the equation of the spectral curve and to the conclusion that each term in the $\hbar$-expansion of~$\bK$ is meromorphic. As a corollary of the determinantal formulas, we prove that the $\omega$-differentials are meromorphic. If the goal were merely to compute $\omega^{(g)}_n$ in closed form as globally meromorphic differentials, we would be done. We wish, however, to obtain a recursion of topological type for these differentials. For that purpose, we analyze the resulting expression for~$\bK$ more closely and derive a list of all its possible poles. As a corollary, we obtain a corresponding list of candidates for the possible poles of the $\omega$-differentials.

In the last step of the proof of topological recursion, in Section~\ref{sec:rho-symmetry}, we need to identify the principal parts of the poles. This is where the theory of $\xy$ swap and symplectic duality comes into play. These are transformations of systems of differentials, and we review them in Appendix~\ref{sec:xy-sympl}. Then, in Section~\ref{sec:rho-symmetry}, we prove the main technical identity: for a suitably chosen weight of the symplectic duality transformation, the original differential is transformed into its image under the involution $\rho\colon (z,w)\mapsto (1/z,w/z^2)$ acting in each argument. The existence of such a transformation has an important consequence: the dual differentials are holomorphic at the zeros of $dX$ and at infinity. This fact is a reformulation of the symplectic duality loop equations at these points. Slightly modified but similar arguments also imply the loop equations at the antidiagonals. This completes the proof of topological recursion.

\subsection{A guide to the Appendices}

In Appendix~\ref{appx:spin-sign} we discuss in detail the conventions for the weights in the spin count of hypermaps. The definition of weights is available in~\cite{EOP,Gun}, but we want to make it more explicit and combinatorial.

Appendix~\ref{sec:xy-sympl} is a review of the theory of $\xy$ swap and symplectic duality. This review is adapted to the case of $\mathbb{Z}_2$-equivariant topological recursion in the sense of~\cite{GKL}, with an odd bi-differential playing the role of a Bergman kernel. The latter theory might be seen as a natural framework for the type of topological recursion on a curve with an involution that we have in this paper. But in fact we argue the opposite: it can be completely incorporated in the realm of a generalized approach to topological recursion. Some new effects emerge from the interaction of the involution and integrability leading to natural statements on global BKP integrability.

In Appendix~\ref{apx:HyperGeom} we recall the setup of hypergeometric BKP tau functions following~\cite{Alexandrov-BGW,GKL,AS23}. In particular, we present a proof of an open conjecture in~\cite{Alexandrov-BGW} on topological recursion for hypergeometric BKP tau functions with rational weights.

\subsection{Acknowledgements}
A.~A. was supported by the Institute for Basic Science (IBS-R003-D1).
P.D.-B. and M.K. were supported by the Basic Research Program of HSE University (HSE-BR-2025-001).
S.~S. was supported by the Dutch Research Council, grant no. OCENW.M.21.233.

The discussions leading to this paper partly took place while some subsets of the authors were visiting the Institut des Hautes Études Scientifiques in Bures-sur-Yvette, the Center for Geometry and Physics of the Institute for Basic Science in Pohang, and the MATRIX programme ``Enumerative Geometry, Integrability, and Topological Recursion'' in Creswick. We thank all these institutions for their hospitality.

\subsubsection{Parallel work of Hock, Lindner and Wulkenhaar}
Some of the important techniques and ideas developed in this paper were independently discovered and developed by A.~Hock, J.~Lindner, and R.~Wulkenhaar in~\cite{Quartic5}, in the context of a completely different application---the so-called quartic Kontsevich model~\cite{Quartic1,Quartic2}, for which an all-genus solution is conjectured in~\cite{Quartic5} using these methods. In particular, this concerns the interaction of the $\xy$ swap and symplectic duality with the $\Z_2$-equivariant version of topological recursion.

In this context, earlier steps towards some of these ideas can already be found in~\cite{Quartic3,Quartic4}, in particular in the treatment of poles at the anti-diagonals and at the fixed point of the group action in the genus $\leq 1$ sectors of the topological expansion.

We thank A.~Hock for informing us about this work.

\subsubsection{AI usage disclosure} The paper is fully designed, thought through, and written by the authors. However, we experimented with AI agents. In particular, for guessing the right structure we let AI agents perform multiple computational experiments (that in the previous epoch  we used to manually set up in packages for symbolic computations) as well as propose proofs of formalizable intermediate steps. While we cannot report that at any of the stages of this project usage of AI was strictly necessary, its usage did speed up our conceptual understanding and did help to polish the final text of the paper.

\section{BKP integrable \texorpdfstring{\(n\)}{n}-point differentials for spin enumeration of
hypermaps}\label{sec:BKP}

\subsection{Schur Q-functions and generating function}

Let polynomials $Q_{a,b}(p)$ be generated by
\begin{equation}\label{eq:genSchur}
\sum_{a,b\geq 0}Q_{a,b} \tau_1^a \tau_2^b=\frac{\tau_1-\tau_2}{\tau_1+\tau_2}\left(\exp\left(2 \sum_{n \ \odd} \frac{p_n}{n} (\tau_1^n+\tau_2^n)\right)-1\right).
\end{equation}
They satisfy $Q_{a,b}=-Q_{b,a}$. 

For a strict partition $\lambda$ we define a {\em Schur Q-function}
\begin{equation}\label{eq:Q-lambda-pfaffian}
Q_\lambda=\Pf_\lambda (Q),
\end{equation}
where on the right-hand side we consider a Pfaffian of the anti-symmetric matrix constructed of the polynomials $Q_{\lambda_i,\lambda_j}$.
We also assume that if $\ell(\lambda)$ is odd, we extend the range of the indices by one, and consider a Pfaffian of the matrix with indices $\{\lambda_1,\dots,\lambda_{\ell(\lambda)},0\}$, see \cite{M} for more details. Note also that $Q_{\emptyset}(p)=1$.

The generating function under consideration is
\begin{equation}
Z=\sum_{\lambda\ \strict}
(2\hbar)^{|\lambda|}
\frac{Q_\lambda(1^{\alpha /(2\hbar)})Q_\lambda(1^{\beta/(2\hbar)})}
     {Q_\lambda(\delta)\,2^{\ell(\lambda)}}Q_\lambda(p).
\label{eq:tau}
\end{equation}
Here \(\lambda\) runs over strict partitions and we consider the following specifications of the Schur Q-functions: $Q_\lambda(\delta)\coloneq \left.Q_\lambda\right|_{p_k=\delta_{k,1}}$, $Q_\lambda(1^{\alpha})\coloneq \left.Q_\lambda\right|_{p_k=\alpha}$, and  $\alpha$ and $\beta$ are two non-vanishing formal parameters.  The corresponding potential is
\begin{equation}
F=\log Z
=\sum_{g\geq0,\,n\geq1}\ \sum_{\mu\ \odd}
\hbar^{2g-2+n}
2^{g-1+n}%
C_g(\mu)\frac{p_\mu}{|\Aut(\mu)|},
\label{eq:potential}
\end{equation}
where $n=\ell(\mu)$.

\begin{remark}\label{rem:power-of-two-convention}
The insertion of powers of~$2$ in the definition of~$Z$ is a matter of convention. Our choice of the rescaling simplifies certain computations with the spectral curve and formulation of the BKP integrability, see below.
\end{remark}

\begin{remark}\label{rem:topological-expansion}
The potential $F=\log Z$ satisfies the topological expansion property: it expands as
\begin{equation}\label{eq:genus-expansion}
F=\sum_{g,n}\hbar^{2g-2+n}F^{(g)}_n,
\end{equation}
where $F^{(g)}_n$ is homogeneous of degree~$n$ and $g$ runs over the nonnegative integers. This statement follows from the combinatorial interpretation of this potential, where $-(2g-2+n)$ is the Euler characteristic of a graph and $g$ is the genus of the surface in which it is embedded. The expansion~\eqref{eq:genus-expansion} can also be derived formally and directly from formula~\eqref{eq:tau} for the exponential of the potential.
\end{remark}

\subsection{BKP integrability}

Generating function \eqref{eq:tau} is a tau function of the BKP hierarchy (more specifically, the BKP hierarchy in this paper is the one that sometimes is referred to as \emph{small} BKP hierarchy).  To prove it we use the lemma about the Schur Q-function expansion of the general solution of the BKP hierarchy, basically following \cite{DJKM}, where the BKP hierarchy was introduced. In this paper we restrict ourselves to the so-called big cell of orthogonal Sato Grassmannian, corresponding to the BKP hierarchy, that is, the tau functions normalized by $Z(0)=1$.

\begin{lemma}[\cite{DJKM}]\label{lemma:BKP}
A function $Z(p)$ is a tau function of the BKP hierarchy if and only if there exists an infinite antisymmetric matrix $A$ whose rows and columns are labelled by nonnegative integers such that the expansion of $Z$ in the basis of the Schur Q-functions is given by
\begin{equation}\label{eq:BKP-Pfaffian-expansion}
Z(p)=\sum_{\lambda\ \strict} \Pf_\lambda(A) Q_\lambda(p),
\end{equation}
where $\Pf_\lambda(A)$ is a Pfaffian for the matrix $A_{\lambda_i,\lambda_j}$ of the size $2{\lceil \frac{\ell(\lambda)}{2}\rceil}$, that is, for odd $\ell(\lambda)$ we consider a Pfaffian of the matrix indexed by $\{\lambda_1,\dots,\lambda_{\ell(\lambda)},0\}$.
\end{lemma}

The components $A_{k,l}$ of $A$ are known as affine coordinates of the corresponding point of the orthogonal Sato Grassmannian.

To establish the BKP integrability of the generating function \eqref{eq:tau} we consider corresponding specifications of the Schur Q-functions in more detail. For $p_k=\delta_{k,1}$ the Schur Q-function is proportional to the Schur Pfaffian \cite[Section III.8]{M},
\begin{equation}\label{eq:Qdel}
Q_\lambda(\delta)=\frac{2^{|\lambda|}}{\prod_{j=1}^{\ell(\lambda)} \lambda_j!}\prod_{i<j} \frac{\lambda_i-\lambda_j}{\lambda_i+\lambda_j}.
\end{equation}

For the second specialization, $p_k=\alpha$, let
\begin{equation}
E_\alpha (\tau_1,\tau_2)
\coloneq \sum_{a,b\geq0}Q_{a,b}(1^\alpha)\tau_1^a\tau_2^b.
\label{eq:E}
\end{equation}
Then from \eqref{eq:genSchur} we have
\begin{equation}\label{eq:E-closed-form}
E_\alpha (\tau_1,\tau_2)=\frac{\tau_1-\tau_2}{\tau_1+\tau_2}
\left(e^{2\alpha (\arctanh\tau_1+\arctanh\tau_2)}-1\right),
\end{equation}
where we apply the identity $\arctanh(\tau)=\tfrac12\log\frac{1+\tau}{1-\tau}$.
Let
\begin{align}\label{eq:f}
f_a(\alpha)&\coloneq \alpha [\tau^a] \left(\frac{1+\tau}{1-\tau}\right)^\alpha \frac{\tau}{1-\tau^2}=\alpha\sum_{j=0}^{a-1}\binom{\alpha-1}{j}\binom{\alpha+a-1-j}{a-1-j},\\
g_a(\alpha)&\coloneq [\tau^a] \left(\frac{1+\tau}{1-\tau}\right)^\alpha \frac{1+\tau^2}{1-\tau^2}=\frac{f_{a+1}(\alpha)+f_{a-1}(\alpha)}{\alpha}.
\label{eq:g}
\end{align}
\begin{lemma}\label{lem:Qab-from-f-and-g}
For $(a,b)\neq (0,0)$
\begin{equation}\label{eq:Qab-from-f-and-g}
Q_{a,b}(1^\alpha)=2 \frac{f_a(\alpha)g_b(\alpha)-g_a(\alpha)f_b(\alpha)}{a+b}.
\end{equation}

\end{lemma}
\begin{proof}
The lemma follows from the identity
\begin{equation}\label{eq:Euler-operator-on-E}
\left(\tau_1\frac{\p}{\p \tau_1}+\tau_2\frac{\p}{\p \tau_2}\right)E_\alpha (\tau_1,\tau_2)=2\alpha\left(\frac{1+\tau_1}{1-\tau_1}\right)^\alpha\left(\frac{1+\tau_2}{1-\tau_2}\right)^\alpha\frac{\tau_1(1+\tau_2^2)-\tau_2(1+\tau_1^2)}{(1-\tau_1^2)(1-\tau_2^2)}
\end{equation}
obtained by a direct computation.
\end{proof}
Therefore
\begin{equation}\label{eq:Q1}
Q_\lambda(1^\alpha)=2^{\lceil \frac{\ell(\lambda)}{2}\rceil}\prod_{j=1}^{\ell(\lambda)}g_{\lambda_j}(\alpha) \Pf \left(\frac{h_{\lambda_i}(\alpha)-h_{\lambda_j}(\alpha)}{\lambda_i+\lambda_j}\right),
\end{equation}
where $h_a(\alpha)=f_a(\alpha)/g_a(\alpha)$.

\begin{theorem}\label{thm:Z-is-BKP}
$Z$ is a BKP tau function.
\end{theorem}

\begin{proof}
The proof is based on the following Pfaffian identity \cite[Equation (14)]{L}:
\begin{equation}\label{eq:Pfaffian-identity-L}
\Pf\left(\frac{(a_i-a_j)(b_i-b_j)}{x_i^2-x_j^2}\right)=\Pf\left(\frac{a_i-a_j}{x_i+x_j}\right)\Pf\left(\frac{b_i-b_j}{x_i+x_j}\right)
\prod_{i<j}\frac{x_i+x_j}{x_i-x_j}.
\end{equation}
Indeed, for the Schur Q-function expansion coefficient of \eqref{eq:tau} from \eqref{eq:Qdel} and \eqref{eq:Q1} one has
\begin{multline}\label{eq:coefficient-as-Pfaffian-product}
(2\hbar)^{|\lambda|}\frac{Q_\lambda(1^{\alpha /(2\hbar)})Q_\lambda(1^{\beta/(2\hbar)})}{Q_\lambda(\delta)\,2^{\ell(\lambda)}}=\hbar^{|\lambda|}2^{2\lceil \frac{\ell(\lambda)}{2}\rceil-\ell(\lambda)}\prod_{j=1}^{\ell(\lambda)}\lambda_j! g_{\lambda_j}\left(\frac{\alpha}{2 \hbar}\right)g_{\lambda_j}\left(\frac{\beta}{2 \hbar}\right)\\
\times\Pf \left(\frac{h_{\lambda_i}\left(\frac{\alpha}{2 \hbar}\right)-h_{\lambda_j}\left(\frac{\alpha}{2 \hbar}\right)}{\lambda_i+\lambda_j}\right)\Pf \left(\frac{h_{\lambda_i}\left(\frac{\beta}{2 \hbar}\right)-h_{\lambda_j}\left(\frac{\beta}{2 \hbar}\right)}{\lambda_i+\lambda_j}\right)\prod_{i<j} \frac{\lambda_i+\lambda_j}{\lambda_i-\lambda_j}.
\end{multline}
This is equal to $\Pf_\lambda(A)$ with
\begin{multline}\label{eq:Aexp}
A_{k,l}=\hbar^{k+l}2^{\delta_{k,0}+\delta_{l,0}}  k!l! \\
\times\frac{\left(f_{k}\left(\frac{\alpha}{2 \hbar}\right)g_{l}\left(\frac{\alpha}{2 \hbar}\right)-g_{k}\left(\frac{\alpha}{2 \hbar}\right)f_{l}\left(\frac{\alpha}{2 \hbar}\right)\right)\left(f_{k}\left(\frac{\beta}{2 \hbar}\right)g_{l}\left(\frac{\beta}{2 \hbar}\right)-g_{k}\left(\frac{\beta}{2 \hbar}\right)f_{l}\left(\frac{\beta}{2 \hbar}\right)\right)}{ k^2-l^2},
\end{multline}
and the statement of the theorem follows from Lemma \ref{lemma:BKP}.
\end{proof}
\begin{remark}\label{rem:MMN-attribution}
BKP integrability of the generating function \eqref{eq:tau} was noted in \cite{MMN}.
\end{remark}
\begin{remark}\label{rem:affine-coordinates}
Expression \eqref{eq:Aexp} together with \eqref{eq:f}, \eqref{eq:g} provides an explicit description of the affine coordinates of the BKP tau function \eqref{eq:tau}.
\end{remark}

\subsection{Kernel}
Let us describe the Baker--Akhiezer kernel that completely specifies the tau function \eqref{eq:tau}.

The Schur \(Q\)-functions satisfy the following properties:
\begin{itemize}
\item \(Q_\lambda|_{p_k=X^k-Y^k}=0\) unless \(\ell(\lambda)\leq2\), and
\begin{equation}
\left.\frac{Q_{a,b}}{2^{\ell(a,b)}}\right|_{p_k=X^k+Y^k}
=\frac{X+Y}{X-Y}(X^aY^b-X^bY^a),
\label{eq:Q-two-row-at-XY}
\end{equation}
where $\ell(a,b)=2$ if both $a>0$ and $b>0$, and $\ell(a,b)=1$ if $ab=0$.

\item \begin{equation}
Q_{a,b}(\delta)
=\frac{a-b}{a+b}\frac{2^{a+b}}{a!b!}.
\label{eq:Q-at-delta}
\end{equation}
\end{itemize}

Define the scalar Baker--Akhiezer kernel
\begin{equation}
K^{\rm tot}(X_1,X_2)=
\left.\frac12\frac{X_1+X_2}{X_1-X_2}
Z\right|_{p_k=X_1^k-X_2^k}.
\label{eq:Kreg}
\end{equation}
Then, substituting the obtained formulas into~\eqref{eq:BKP-Pfaffian-expansion} and \eqref{eq:tau} we obtain an expansion
\begin{align}
K^{\rm tot}(X_1,X_2)
={}&\frac{1}{2}\sum_{a>b\ge0}2^{\ell(a,b)}A_{a,b}\bigl(X_1^a(-X_2)^b-X_1^b(-X_2)^a\bigr)
+\frac12\frac{X_1+X_2}{X_1-X_2}\label{eq:Kregsum-affine-coordinates}
\\={}&\frac12\sum_{a>b\geq0}
\hbar^{a+b}\bigl(X_1^a(-X_2)^b-X_1^b(-X_2)^a\bigr)
\frac{a+b}{a-b}a!b!
\nonumber\\[-0.2em]
&\qquad\times
\coeff{\sigma_1^a\sigma_2}{b}E_{\alpha /(2\hbar)}(\sigma_1,-\sigma_2)
\coeff{\tau_1^a\tau_2}{b}E_{\beta/(2\hbar)}(\tau_1,-\tau_2)
+\frac12\frac{X_1+X_2}{X_1-X_2}.
\label{eq:Kregsum}
\end{align}

It is useful to introduce
\begin{equation}
\widehat E_\alpha (\tau_1,\tau_2)\coloneq
\frac{1-\tau_1\tau_2}{1+\tau_1\tau_2}
e^{2\alpha (\arctanh\tau_1+\arctanh\tau_2)}.
\label{eq:Ehat}
\end{equation}
By a direct substitution, one checks the identity
\begin{equation}\label{eq:E-Ehat-derivative}
 (\tau_1\partial_{\tau_1}+\tau_2\partial_{\tau_2})\;E_\alpha(\tau_1,\tau_2)=
 (\tau_1\partial_{\tau_1}-\tau_2\partial_{\tau_2})\;\widehat E_\alpha(\tau_1,\tau_2).
\end{equation}
 For \(a\neq b\), it implies
\begin{equation}
\frac{a+b}{a-b}Q_{a,b}(1^\alpha )
=\coeff{\tau_1^a\tau_2}{b}\widehat E_\alpha (\tau_1,\tau_2).
\label{eq:QEhat}
\end{equation}
Although the left-hand side is undefined for \(a=b\), the coefficient on the
right-hand side is well defined. In the symmetrized expression below, its diagonal
contribution is multiplied by the vanishing diagonal coefficient of \(E\).
Consequently,
\begin{align}
K^{\rm tot}(X_1,X_2)
={}&\frac12\sum_{a,b\geq0}
\hbar^{a+b}X_1^a(-X_2)^b\;a!b!
\nonumber\\[-0.2em]
&\quad\times
\coeff{\sigma_1^a\tau_1^a\sigma_2^b\tau_2}{b}
E_{\alpha /(2\hbar)}(\sigma_1,-\sigma_2)
\widehat E_{\beta/(2\hbar)}(\tau_1,-\tau_2)+\frac12\frac{X_1+X_2}{X_1-X_2}.
\label{eq:Kregsym}
\end{align}


\subsection{BKP and determinantal identities}
The $n$-point functions and differentials are defined by
\begin{align}
H_{g,n}(X_1,\ldots,X_n)
&=\left.\coeff{\hbar}{2g-2+n}
\prod_{i=1}^{n}
\left(\sum_{\substack{k\in\mathbb Z_{>0}\\ k\ \odd}}
X_i^k\frac{\partial}{\partial p_k}\right)F\right|_{p=0},
\label{eq:Hgn}\allowdisplaybreaks[1]\\
&=2^{g-1+n}\sum_{\substack{k_1,\ldots,k_n>0\\ k_i\ \odd}}
C_g(k_1,\ldots,k_n)X_1^{k_1}\cdots X_n^{k_n},
\label{eq:Hgnexp}
\end{align}
and
\begin{equation}
\omega^{(g)}_n=d_1\cdots d_n H_{g,n}
+\delta_{g,0}\delta_{n,2}
\left(\frac{d X_1d X_2}{(X_1-X_2)^2}
+\frac{d X_1d X_2}{(X_1+X_2)^2}\right).
\label{eq:omega}
\end{equation}

Introduce a bi-half-differential version of the Baker--Akhiezer kernel for the BKP tau function~\eqref{eq:tau}:
\begin{equation}
\bK^{\rm tot}(X_1,X_2)=K^{\rm tot}(X_1,X_2)\sqrt{\frac{d X_1}{X_1}}\sqrt{\frac{d X_2}{X_2}}
=\left.\frac12\frac{X_1+X_2}{X_1-X_2}
\sqrt{\frac{d X_1}{X_1}}\sqrt{\frac{d X_2}{X_2}}
Z(p)\right|_{p_k=X_1^k-X_2^k}.
\label{eq:BA}
\end{equation}

Introduce the antisymmetrization operators
\begin{equation}
({\rm Alt}_i f)(X_i)=f(X_i)-f(-X_i).
\label{eq:antisym}
\end{equation}
For
\begin{equation}
\omega^{\rm tot}_n=\sum_{g\geq0}\hbar^{2g-2+n}\omega^{(g)}_n,
\label{eq:omegan}
\end{equation}
the BKP integrability implies the determinantal identities, derived in \cite{WY}, see also Appendix~\ref{sec:xy-sympl}:
\begin{equation}
\omega^{\rm tot}_1(X_1)=
\left.\left(
\bK^{\rm tot}(X_1,X_2)-\frac12\frac{X_1+X_2}{X_1-X_2}
\sqrt{\frac{d X_1}{X_1}}\sqrt{\frac{d X_2}{X_2}}
\right)\right|_{X_2=X_1},
\label{eq:omega1}
\end{equation}
and, for \(n\geq2\),
\begin{equation}
\omega^{\rm tot}_n={\rm Alt}_1\cdots{\rm Alt}_{n-1}(-1)^{n-1}
\sum_{\sigma\in\operatorname{cycl}(n)}\prod_{i=1}^{n}\bK^{\rm tot}(X_i,X_{\sigma(i)}).
\label{eq:detid}
\end{equation}
One may also antisymmetrize in the last variable \(X_n\), with an additional
factor \(1/2\), but this is unnecessary: skew-symmetry in \(X_n\) follows
from
\begin{equation}
\bK^{\rm tot}(-X_1,-X_2)=-\bK^{\rm tot}(X_2,X_1),
\label{eq:BKPksym}
\end{equation}
which is valid for all BKP kernels.  The use of bi-half-differentials makes
the kernel coordinate invariant and allows the determinantal identities to be
globalized on the spectral curve.

\subsection{The stabilizing}

For many computations, it is more convenient to use the ``stabilized'' versions of $Z$ and $F$ obtained by dropping the most ``unstable'' contribution of the $(0,1)$ terms:
\begin{align}
\widetilde F&=F-\hbar^{-1}F^{(0)}_1,\qquad \widetilde Z=e^{-\hbar^{-1}F^{(0)}_1}Z,\label{eq:rescaled-F-Z}
\\
\omega_n&=\omega^{\rm tot}_n-\hbar^{-1}\delta_{n,1}\omega^{(0)}_1
=\sum_{\substack{g\ge0\\(g,n)\ne(0,1)}}\hbar^{2g-2+n}\omega^{(g)}_n,\label{eq:rescaled-omega}
\\
\bK(X_1,X_2)&=
\left.\frac12\frac{X_1+X_2}{X_1-X_2}
\sqrt{\frac{d X_1}{X_1}}\sqrt{\frac{d X_2}{X_2}}
\;\widetilde Z\right|_{p_k=X_1^k-X_2^k}\label{eq:rescaled-K-def}
\\
&=e^{-\hbar^{-1}\bigl(H^{(0)}_1(X_1)-H^{(0)}_1(X_2)\bigr)}\bK^{\rm tot}(X_1,X_2).
\label{eq:rescaled-K-vs-Ktot}
\end{align}
Changing the linear terms in the potential preserves BKP integrability, and the determinantal identities are preserved under the stabilization:
\begin{align}
\omega_1&=
\left(
\bK(X_1,X_2)-\frac12\frac{X_1+X_2}{X_1-X_2}
\sqrt{\frac{d X_1}{X_1}}\sqrt{\frac{d X_2}{X_2}}\right)\Big|_{X_2=X_1},\label{eq:omega1-rescaled}
\\
\omega_n&={\rm Alt}_1\cdots{\rm Alt}_{n-1}(-1)^{n-1}
\sum_{\sigma\in\operatorname{cycl}(n)}\prod_{i=1}^{n}\bK(X_i,X_{\sigma(i)}),
\quad n\ge2.\label{eq:detid-rescaled}
\end{align}

The ``stabilized'' objects have the advantage of being expandable in nonnegative powers of~$\hbar$, whereas the original ones can only be treated as transseries. For this reason, throughout the computations in this paper we reserve the notation~$\bK$, $\omega_n$ for the corresponding ``stabilized'' versions of these objects.

%


 \section{Integral representations and spectral curve} \label{sec:K-Gauss}

\subsection{Formal Gaussian integral representation}

Let $Q$ be a nondegenerate symmetric matrix representing the nondegenerate quadratic form $Q(\tau)=\tau^\top Q\tau$, $\tau\in\C^n$. Denote its inverse by $C=Q^{-1}$; this is the \emph{covariance}. Recall that a general formal $n$-fold Gaussian integral is a linear functional on the space of polynomials in $n$ variables defined by the formal rule called the \emph{Wick expansion formula} (see, e.g.,~\cite{Ng} or the very beginning of~\cite{ZinnJustin-book}):
\begin{equation}\label{eq:wick-expansion}
\begin{aligned}
I(\cA)&=\left(\frac{1}{2\pi}\right)^{n/2}\int\dots\int\cA(\tau_1,\dots,\tau_n)\;e^{-\frac12Q(\tau)}\dd\tau_1\dots\dd\tau_n
\\&:=\tfrac{1}{\sqrt{\det Q}}\;e^{\frac12\sum_{i,j=1}^n C_{ij}\frac{\partial^2}{\partial\tau_i\partial\tau_j}}\cA(\tau)\bigm|_{\tau=0}.
\end{aligned}
\end{equation}
The formula itself is given on the second line, while the integral expression on the first line serves merely as notation for this expansion. If $Q$ is real and positive definite, it coincides with an ordinary analytic integral over $\R^n$, but we shall use it formally without any additional assumption on $Q$. The choice of sign of the square root of the determinant is an ambiguity corresponding to a ``choice of orientation of the integration cycle.'' In applications, the function $\cA$ called \emph{amplitude} is expanded as a series in an additional formal variable, with coefficients polynomial in $\tau$. The integral is then also a formal series in that variable.

The Gaussian integral shares most of the properties of ordinary integrals, such as additivity and Fubini factorization. The strongest is its invariance under changes of variables, with the usual Jacobian factor. The allowed changes of variables are invertible linear transformations plus series in positive powers of an additional parameter~$\epsilon$ with polynomial coefficients. Also, the definition~\eqref{eq:wick-expansion} makes sense if the quadratic form $Q$ depends on $\epsilon$ such that $Q=Q_0+O(\epsilon)$ where~$Q_0$ is nondegenerate. Then $Q$ can be replaced by~$Q_0$ in the Gaussian exponential term, with the expansion of the remaining part $Q-Q_0$ accounted for in the corresponding modification of the amplitude. This transformation also preserves the integral.

Another useful property is the integration-by-parts identity, or Stokes' formula: the integral of the differential of an $(n-1)$-form vanishes.

\bigskip
As an example, consider the expression in Eq.~\eqref{eq:Kregsym}. It involves the functional
\begin{equation}
\mathcal I_c(f)=\sum_{a\geq0}c^a a!
\coeff{\sigma^a{\tau}}{a}f(\sigma,\tau)=e^{c \frac{\p^2}{\p \sigma \p \tau}}f(\sigma,\tau)\bigm|_{\sigma=\tau=0}.
\label{eq:functional}
\end{equation}
This functional has the formal Gaussian representation
\begin{equation}
\mathcal I_c(f)=\frac{i}{2\pi c}\iint
f(\sigma,\tau)e^{-\frac{\sigma\tau}{c}}\dd\sigma\dd\tau.
\label{eq:formalGaussian}
\end{equation}
Define the \emph{amplitude}
\begin{equation}
\mathcal A(\sigma_1,\tau_1,\sigma_2,\tau_2)=
\frac{\sigma_1+\sigma_2}{\sigma_1-\sigma_2}
\frac{1+\tau_1\tau_2}{1-\tau_1\tau_2},
\label{eq:amplitude}
\end{equation}
and the \emph{phase}
\begin{equation}
\Phi_X(\sigma,\tau)=-\frac{\sigma\tau}{X}
+\alpha \arctanh(\sigma)+\beta\arctanh(\tau).
\label{eq:phase}
\end{equation}
Then, in the same way, we obtain from~\eqref{eq:Kregsym}:
\begin{equation}
\begin{aligned}\label{eq:fourGaussian}
K^{\rm tot}(X_1,X_2)&
=\frac12\frac{X_1+X_2}{X_1-X_2}-\frac{1}{2}\frac{1}{X_1X_2}
\left(\frac{i}{2\pi\hbar}\right)^2
\iiiint E(\sigma_1,-\sigma_2)\,\widehat E(\tau_1,-\tau_2)\,
e^{-\frac{\sigma_1\tau_1}{\hbar X_1}}
\dd\sigma_1\dd\tau_1
e^{\frac{\sigma_2\tau_2}{\hbar X_2}}
\dd\sigma_2\dd\tau_2
\\
&=-\frac{1}{2}\frac{1}{X_1X_2}
\left(\frac{i}{2\pi\hbar}\right)^2
\iiiint\mathcal A(\sigma_1,\tau_1,\sigma_2,\tau_2)
e^{\frac1\hbar\Phi_{X_1}(\sigma_1,\tau_1)}
\dd\sigma_1\dd\tau_1
e^{-\frac1\hbar\Phi_{X_2}(\sigma_2,\tau_2)}
\dd\sigma_2\dd\tau_2.
\end{aligned}
\end{equation}

The equality on the last line contains implicitly two assertions. First, it assumes a certain framework for which this integral with singular amplitude makes sense, and second, it claims that the Cauchy singular term can be represented within this framework by the integral
\begin{multline}\label{eq:Cauchy}
\frac{1}{2}\frac{X_1+X_2}{X_1-X_2}=-\frac{1}{2\,X_1X_2}\Bigl(\frac{i}{2\pi\hbar}\Bigr)^2\iiiint
\frac{\sigma_1+\sigma_2}{\sigma_1-\sigma_2}
\frac{1+\tau_1\tau_2}{1-\tau_1\tau_2}
\\\times e^{\frac{1}{\hbar}\left(-\frac{\sigma_1\tau_1}{X_1}+\frac{\sigma_2\tau_2}{X_2}
+\beta\,\arctanh(\tau_1)-\beta\,\arctanh(\tau_2)\right)}d\sigma_1d\tau_1d\sigma_2d\tau_2
\end{multline}

In order to justify~\eqref{eq:fourGaussian}, one can insert artificially an additional formal parameter~$\epsilon$ into all integrals under consideration and consider them in the formal expansion with respect to this parameter. For example, one can apply the rescaling of parameters
\begin{equation}\label{eq:Xst-scaling}
X_i\to\epsilon X_i,\quad \hbar\to\epsilon^2\hbar,\quad \alpha\to\epsilon^2\alpha,\quad\beta\to\beta,
\end{equation}
and also the corresponding rescaling of the integration variables defined by
\begin{equation}\label{eq:sigmatau-scaling}
\sigma_i=\epsilon\,\beta\, X_i+\epsilon^2\xi_i,\qquad
\tau_i=\epsilon\,\eta_i.
\end{equation}
With this rescaling, all three integrals in~\eqref{eq:fourGaussian}--\eqref{eq:Cauchy} take the standard form with quadratic Gaussian exponential and with the amplitude polynomial in $\xi_1,\eta_1,\xi_2,\eta_2$ in each order in~$\epsilon$, so that they can be compared term-by-term in each order of $\epsilon$-expansion. Moreover, the first integral in~\eqref{eq:fourGaussian} has regular amplitude and can be reexpanded as a series in~$X_1,X_2$ with coefficients depending on $\epsilon$ in a polynomial way, hence it admits a specialization $\epsilon=1$ recovering the original definition~\eqref{eq:Kregsym} of~$K^{\rm tot}$.
 
Now, let us prove~\eqref{eq:Cauchy}. In order to simplify arguments we will use the original coordinates; the accepted $\epsilon$-scaling makes all changes in the computations below admissible formal changes of coordinates. Let us start with the change of variables
\begin{equation}
\rho_i=(1-\tau_1\tau_2)\,\sigma_i,\quad\zeta_i=\frac{\tau_i}{1-\tau_1\tau_2}.
\end{equation}
Then, by the equalities
\begin{gather}
\arctanh(\tau_1)-\arctanh(\tau_2)=\arctanh(\zeta_1-\zeta_2),
\\
\frac{\sigma_1+\sigma_2}{\sigma_1-\sigma_2}
=\frac{\rho_1+\rho_2}{\rho_1-\rho_2},
\\
\sigma_i\tau_i=\rho_i\zeta_i,
\\
\det\frac{\partial(\rho_1,\zeta_1,\rho_2,\zeta_2)}{\partial(\sigma_1,\tau_1,\sigma_2,\tau_2)}
=\frac{1+\tau_1\tau_2}{1-\tau_1\tau_2}
\end{gather}
we bring the integral on the right-hand side of~\eqref{eq:Cauchy} to the form
\begin{equation}
-\frac{1}{2\,X_1X_2}\Bigl(\frac{i}{2\pi\hbar}\Bigr)^2\iiiint
\frac{\rho_1+\rho_2}{\rho_1-\rho_2}
e^{\frac{1}{\hbar}\left(-\frac{\rho_1\zeta_1}{X_1}+\frac{\rho_2\zeta_2}{X_2}
+\beta\,\arctanh(\zeta_1-\zeta_2)\right)}d\rho_1d\zeta_1d\rho_2d\zeta_2.
\end{equation}
Next, applying one more change by passing to the variables
\begin{equation}
a=\frac{\rho_1}{X_1}-\beta,
\quad b=\frac{\rho_1}{X_1}-\frac{\rho_2}{X_2},
\quad u=\zeta_1-\zeta_2,\quad v=\zeta_2
\end{equation}
we rewrite the integral finally as
\begin{equation}
\frac{1}{2}\Bigl(\frac{i}{2\pi\hbar}\Bigr)^2\iiiint
\frac{(X_1+X_2)\,(a+\beta)-X_2b}{(X_1-X_2)\,(a+\beta)+X_2b}
e^{\frac{\beta}{\hbar}\,(\arctanh(u)-u)}e^{-\frac{1}{\hbar}(au+bv)}da\,du\,db\,dv.
\end{equation}
Since the amplitude is independent of~$v$, the integration over $(b,v)$ simply sets $b=0$. Since the resulting amplitude
\begin{equation}
\frac{(X_1+X_2)\,(a+\beta)-X_2b}{(X_1-X_2)\,(a+\beta)+X_2b}
e^{\frac{\beta}{\hbar}\,(\arctanh(u)-u)}\Bigm|_{b=0}=
\frac{X_1+X_2}{X_1-X_2}
e^{\frac{\beta}{\hbar}\,(\arctanh(u)-u)}
\end{equation}
is independent of~$a$ we compute the resulting integral over $(a,u)$ by setting $u=0$ in the amplitude that gives exactly $\frac12\frac{X_1+X_2}{X_1-X_2}$. This completes the proof of~\eqref{eq:Cauchy} and~\eqref{eq:fourGaussian}.\qed


\begin{remark}\label{rem:matrix-integral-conjecture}
Integral representation \eqref{eq:fourGaussian} hints that the tau function for the spin hypermaps \eqref{eq:tau} can be described by a matrix integral
of Kontsevich type, similar to the one for the simple singularity of type D derived in~\cite{AM}. We conjecture that a natural matrix integral description exists, however, do not discuss it here. Moreover, for specific values of $\alpha$ and $\beta$
one can expect existence of the analogs of the two-matrix and complex matrix integral descriptions for the original hypermaps.
\end{remark}

\begin{remark}\label{rem:twin-equivalence}
It is possible to exchange the roles of $E$ and $\widehat E$ (or, equivalently, to exchange the roles of $\alpha $ and~$\beta$) in the expansion~\eqref{eq:Kregsym}. This would lead to a different but equivalent representation of the kernel as a Gaussian integral:
\begin{align}
\cA^{\rm tw}(\sigma_1,\tau_1,\sigma_2,\tau_2)&=
\frac{1+\sigma_1\sigma_2}{1-\sigma_1\sigma_2}\frac{\tau_1+\tau_2}{\tau_1-\tau_2},\label{eq:amplitude-twin}
\\
K^{\rm tot}(X_1,X_2)&=-\frac{1}{2}\frac{1}{X_1X_2}
\left(\frac{i}{2\pi\hbar}\right)^2
\iiiint\mathcal A^{\rm tw}(\sigma_1,\tau_1,\sigma_2,\tau_2)
\notag\\&\qquad\times
e^{\frac1\hbar\Phi_{X_1}(\sigma_1,\tau_1)}
\dd\sigma_1\dd\tau_1
e^{-\frac1\hbar\Phi_{X_2}(\sigma_2,\tau_2)}
\dd\sigma_2\dd\tau_2.\label{eq:fourGaussian-twin}
\end{align}
\end{remark}

\subsection{Wave function and quantum curve}

Let us consider the wave function for the BKP tau function $Z$:
\begin{equation}
\Psi(X)\coloneq \left.Z(p)\right|_{p_k=X^k}.\label{eq:wave-function}
\end{equation}
It coincides with a specialization of the Baker--Akhiezer kernel \eqref{eq:Kreg}, $\Psi(X)=2 K^{\rm tot}(X,0)$ and can be described by an integral,
\begin{equation}
\Psi(X)=\frac{i}{2\pi \hbar X}\iint e^{\frac1\hbar\Phi_{X}(\sigma ,\tau)}
\dd\sigma \dd\tau.\label{eq:wave-function-integral}
\end{equation}
Let $D\coloneq  \hbar X \frac{\p}{\p X}$. The following statement can be proven with integration by parts:
\begin{proposition}\label{prop:quantum-curve}
The wave function satisfies the quantum spectral curve equation
\begin{align}
& \Bigl(\bigl(1-X^2D\,(D-\hbar)\bigr)\,\bigl(1-X^2D\,(D+\hbar)\bigr)\,D\Bigr.\qquad\label{eq:quantum-curve}
\\ \notag &
  \qquad -\Bigl.(\alpha^2+\beta^2)\,X^2D-\alpha\beta \,X\,\bigl(1+X^2D\,(D+\hbar)\bigr)\Bigr)\,\Psi=0.
\end{align}
\end{proposition}
In the semi-classical limit, if one substitutes $D$ with $y$ and sets $\hbar=0$, the spectral curve reduces to
\begin{equation}
y\left(1-X^2y^2\right)^2-(\alpha^2+\beta^2)X^2y- \alpha\beta X(1+X^2y^2) =0,\label{eq:quantum-curve-classical-limit}
\end{equation}
which is equivalent to \eqref{eq:spectral-curve-intro}, \eqref{eq:X-in-z-intro} with $y=z/X$.

\subsection{Stationary phase analysis}

The standard treatment of integrals of the form~\eqref{eq:fourGaussian} is to expand them near a critical point of the phase. Consider one factor of~\eqref{eq:fourGaussian}:
\begin{equation}
I(X,\hbar)=\frac{1}{X}\frac{i}{2\pi\hbar}
\iint F(\sigma,\tau)e^{\frac1\hbar\Phi_X(\sigma,\tau)}
\dd\sigma\dd\tau,
\label{eq:singleIntegral}
\end{equation}
for some function (a power series) \(F\). The phase has a critical point $(\sigma,\tau)=(s,t)$ near the origin whose position depends on~$X$. The critical point equations for \(\Phi_X\) are
\begin{equation}
\alpha X=t(1-s^2),\qquad \beta X=s(1-{t}^2).
\label{eq:critical}
\end{equation}
These equations define implicit functions $s=s(X)$ and $t=t(X)$ with the branches near the origin characterized by
\(s(X)=\beta X+O(X^3)\) and \(t(X)=\alpha X+O(X^3)\).  Set
\begin{equation}
H^{(0)}_1(X)=\Phi_X(s(X),t(X)).
\label{eq:H01def}
\end{equation}
Explicitly,
\begin{equation}
H^{(0)}_1(X)=\alpha \arctanh s(X)+\beta\arctanh t(X)-\frac{s(X)t(X)}{X}.
\label{eq:H01}
\end{equation}
Introduce the shifted phase
\begin{equation}\label{eq:shifted-phase}
W_X(\sigma,\tau)=\Phi_X(\sigma,\tau)-H^{(0)}_1(X).
\end{equation}

\begin{proposition}\label{prop:hbar-1G}
Assume that $F(\sigma,\tau)$ is a rational function regular at the origin. Then, the stabilized integral
\begin{equation}
\widetilde I(X,\hbar)=e^{-\frac1\hbar H^{(0)}_1(X)}I(X,\hbar)
=\frac{1}{X}\frac{i}{2\pi\hbar}
\iint F(\sigma,\tau)
e^{\frac1\hbar W_X(\sigma,\tau)}
\dd\sigma\dd\tau
\label{eq:rescaledSingle}
\end{equation}
has an expansion in nonnegative powers of \(\hbar\).
\end{proposition}

\begin{proof} Apply the following change of the integration variables:
\begin{equation}\label{eq:gaussian-shift-substitution}
\sigma=s+\sqrt{\hbar}\,\chi,\quad \tau=t+\sqrt{\hbar}\,\eta,
\end{equation}
where $s=s(X)$ and $t=t(X)$ correspond to the critical point of $W_X$. Since $s(0)=t(0)=0$, this is an allowed change with $X$ treated as a formal small parameter. The expansion of~$W_X/\hbar$ in $\chi$ and $\eta$ has no constant or linear terms, while the quadratic terms are independent of $\hbar$ and are given by the matrix
\begin{equation}
Q=-\dd^2 W_X\big|_{s,t}=\frac{1}{X}
\begin{pmatrix}
-\dfrac{2st}{1-s^2}&1\\[2mm]
1&-\dfrac{2st}{1-{t}^2}
\end{pmatrix},
\label{eq:Hessian}
\end{equation}
with the corresponding covariance
\begin{equation}
C=Q^{-1}
=\frac{X}{\Delta}
\begin{pmatrix}
\dfrac{2\,st}{1-{t}^2}&1\\[2mm]
1&\dfrac{2\,st}{1-s^2}
\end{pmatrix},
\label{eq:covariance}
\end{equation}
where
\begin{equation}
\Delta=1-\frac{4s^2{t}^2}{(1-s^2)(1-{t}^2)},\qquad 
\qquad
\det Q=-\frac{\Delta}{X^2}.
\end{equation}
The higher-order terms contain positive powers of $\hbar$:
\begin{equation}\label{eq:higher-order-phase}
\hbar^{-1}W_X+\tfrac12Q(\chi,\eta)=\sum_{r\ge3}\hbar^{\frac{r-2}{2}}
\left(\frac{\alpha }{r!}\arctanh^{(r)}(s)\,\chi^r+
\frac{\beta}{r!}\arctanh^{(r)}(t)\,\eta^r\right).
\end{equation}
Similarly, the amplitude $F(s+\sqrt\hbar\,\chi,t+\sqrt\hbar\,\eta)$ expands in nonnegative powers of $\sqrt\hbar$, with coefficients polynomial in $\chi,\eta$. We thus obtain the Gaussian integral in its conventional form, with Gaussian exponential $e^{-\frac12Q(\chi,\eta)}$ and the amplitude polynomial at each order of~$\sqrt{\hbar}$ regarded as a small formal parameter. Notice that odd powers of $\sqrt\hbar$ in the amplitude are always accompanied by monomials of odd degree in $\chi,\eta$ and therefore do not contribute to the integral. Consequently, the expansion contains only nonnegative integer powers of~$\hbar$.
\end{proof}

For the kernel, the algebraically consistent stabilization is
\begin{multline}
K(X_1,X_2)=e^{-\frac1\hbar(H^{(0)}_1(X_1)-H^{(0)}_1(X_2))}K^{\rm tot}(X_1,X_2)
\\=-\frac{1}{2X_1X_2}
\left(\frac{i}{2\pi\hbar}\right)^2
\iiiint\mathcal A(\sigma_1,\tau_1,\sigma_2,\tau_2)
\\
\times
e^{\frac1\hbar W_{X_1}(\sigma_1,\tau_1)}
\dd\sigma_1\dd\tau_1
e^{-\frac1\hbar W_{X_2}(\sigma_2,\tau_2)}
\dd\sigma_2\dd\tau_2
\label{eq:rescaledKernel}
\end{multline}
which expands in nonnegative powers of $\hbar$. Comparing this with the definition~\eqref{eq:Kreg} of $K^{\rm tot}$, we see that $H^{(0)}_1$ given by~\eqref{eq:H01} is exactly the genus-zero one-point function and that $K$ is the stabilized version of the $K$-kernel.

\begin{remark}\label{rem:epsilon-rescaling}
Strictly speaking, the arguments in the proof of Proposition~\ref{prop:hbar-1G} require that the amplitude $F$ is a regular series in $\sigma,\tau$ variables. On the other hand, the amplitude in the integral representation~\eqref{eq:fourGaussian} has a pole on the diagonal~$\sigma_1=\sigma_2$ and the equality~\eqref{eq:rescaledKernel} requires an additional justification. This is done by a suitable rescaling of parameters as in~\eqref{eq:Xst-scaling}-\eqref{eq:sigmatau-scaling}. The idea is essentially the same: the comparison with the stationary-phase expansion is obtained by translating to the formal critical point and incorporating the quadratic correction into the Gaussian quadratic form, using a suitable change of variables. Specialization at $\epsilon=1$ is understood coefficientwise in the local $X_1,X_2$-expansion, after clearing the finite diagonal pole at each $\hbar$-order. We skip the technical details since they are standard for the Gaussian-type integrals technology, see e.g.,~\cite{Ng}.
\end{remark}

\subsection{The spectral curve}

The functions $s(X)$ and $t(X)$ defined by~\eqref{eq:critical} were introduced above as germs, or power series, at the origin. They are, however, algebraic.

\begin{definition}\label{def:spectral-curve}
The \emph{spectral curve} is the smooth algebraic curve given in the affine chart by
\eqref{eq:critical}.
\end{definition}
Below we assume that $\alpha$ and $\beta$ satisfy the nondegeneracy condition
\begin{equation}
\alpha\,\beta\,(\alpha^2-\beta^2)\ne0.\label{eq:nondegeneracy-condition}
\end{equation}
Then, eliminating $X$ from the equations of the curve gives a smooth plane cubic, that is, the spectral curve is elliptic.  Put
\begin{equation}
z=s t,\qquad
w=(\beta+\alpha z)t=(\alpha +\beta z)s.
\label{eq:zw}
\end{equation}
Then the curve takes in these coordinates the standard hyperelliptic form
\begin{equation}
w^2=z(\alpha +\beta z)(\beta+\alpha z),
\label{eq:spectralCurve}
\end{equation}
with the elliptic involution \(\iota(z,w)=(z,-w)\). Its function field consists of
expressions \(R_0(z)+R_1(z)w\), with \(R_0,R_1\) rational.  The representation of the basic
functions in the $(z,w)$ coordinates is
\begin{equation}
X=\frac{(1-z^2)w}{(\alpha z+\beta)(\alpha +\beta z)},\qquad
s=\frac{w}{\alpha +\beta z},\qquad
t=\frac{w}{\beta+\alpha z}.
\label{eq:basicfunctions}
\end{equation}

That gives also
\begin{align}\label{eq:R}
\Delta&=\frac{R(z)}
       {(1-z^2)^2},\qquad
R(z)=1-6z^2-4\tfrac{\alpha ^2+\beta^2}{\alpha\beta }z^3-3z^4.
\end{align}


\begin{proposition}\label{prop:rational-expansion}
Assume that the function $F(\sigma,\tau)$ is rational and regular at the origin. Then the integral in Proposition~\ref{prop:hbar-1G} admits an expansion of the form
\begin{equation}\label{eq:2Gintrat}
\frac{1}{X}\frac{i}{2\pi\hbar}
\iint F(\sigma,\tau)
e^{\frac1\hbar W_X(\sigma,\tau)}
\dd\sigma\dd\tau
=\frac{1}{\Delta^{1/2}}\sum_{k=0}^{\infty}B_k(s,t)\hbar^k
\end{equation}
where each $B_k$ is a rational expression in $s,t$, and hence a globally meromorphic function on the spectral curve.
\end{proposition}

\begin{proof} The proof is contained implicitly in the computations in the proof of Proposition~\ref{prop:hbar-1G}. Just notice that all coefficients in the Taylor expansion~\eqref{eq:higher-order-phase} of the phase, all Taylor coefficients of the amplitude $F(s+\sqrt\hbar\,\chi,t+\sqrt\hbar\,\eta)$, and all components of the covariance~\eqref{eq:covariance} are rational in~$s$ and~$t$, that is, globally meromorphic on the spectral curve.
\end{proof}

As a corollary of Proposition~\ref{prop:rational-expansion}, we obtain a similar expansion for the stabilized kernel $K$.

\begin{corollary}\label{cor:BArat}
The stabilized Baker--Akhiezer kernel~\eqref{eq:rescaledKernel} admits an expansion of the form
\begin{equation}\label{eq:BArat}
K(X_1,X_2)=\frac{1}{\Delta_1^{1/2}\Delta_2^{1/2}}
\sum_{k=0}^\infty B_k(s_1,t_1,s_2,t_2)\,\hbar^{k}
\end{equation}
with the coefficients $B_k$ rational in all arguments, that is, meromorphic on the Cartesian square of the spectral curve.
\end{corollary}

It is straightforward to compute
\begin{equation}\label{eq:dX-over-X-normalized}
\frac{1}{\Delta}\frac{dX}{X}=\frac{\alpha\beta}2\,\frac{(1-z^2)}{w^2}\,dz.
\end{equation}
Respectively, we get
\begin{multline}\label{eq:bK-in-z}
\bK(X_1,X_2)=K(X_1,X_2)\sqrt{\frac{dX_1}{X_1}}\sqrt{\frac{dX_2}{X_2}}
=
\frac{\alpha\beta}{2}\biggl(\sum_{k=0}^\infty B_k(s_1,t_1,s_2,t_2)\,\hbar^{k}\biggr)
\prod_{i=1}^2\frac{(1-z_i^2)^{1/2}}{w_i}\,\sqrt{dz_i}.
\end{multline}
\begin{corollary}\label{cor:omegarat}
All differentials $\omega^{(g)}_n$ extend as globally meromorphic differentials on the Cartesian power of the spectral curve.
\end{corollary}

\begin{proof}
Each factor $(1-z_i^2)^{1/2}\,\sqrt{dz_i}$ appears twice in every summand of the determinantal expression. Hence all radicals disappear from the final expression for~$\omega^{(g)}_n$.
\end{proof}

\begin{remark}\label{rem:points-notation}
In view of Corollaries~\ref{cor:BArat} and~\ref{cor:omegarat}, from now on we use points $p_1,p_2,\dots$ of the spectral curve as the arguments of all the (half-)differentials under consideration, and we write $X_i=X(p_i)$, $z_i=z(p_i)$, $w_i=w(p_i)$, and so on.
\end{remark}

\begin{remark}\label{rem:iota-oddness}
The involution $X\mapsto -X$ involved in the formulation of BKP integrability extends to the hyperelliptic involution $\iota:(z,w)\mapsto(z,-w)$ of the spectral curve. It follows that all differentials $\omega^{(g)}_n$ are odd with respect to $\iota^*$ in each argument. In other words, they can be represented in the form
\begin{equation}\label{eq:omega-W-eta}
\omega^{(g)}_n=W^{(g)}_n(z_1,\dots,z_n)\;\eta_1\dots\eta_n,
\end{equation}
where $W^{(g)}_n$ is rational and $\eta=\frac{d z}{w}$ is the basic holomorphic differential.
\end{remark}

\begin{example}\label{ex:omega01-explicit} The arguments in the above proof do not cover an exceptional case $(g,n)=(0,1)$. Meanwhile, the assertion of Corollary~\ref{cor:omegarat} holds true for this case as well. Namely, applying directly~\eqref{eq:H01} and~\eqref{eq:zw}--\eqref{eq:basicfunctions} we compute, explicitly,
\begin{equation}\label{eq:omega01-in-z}
\omega^{(0)}_1=d H^{(0)}_1=y\,\frac{d X}{X},\qquad y=\frac{z}{X}=\frac{w}{1-z^2},
\end{equation}
which agrees with~\eqref{eq:omega01}.
\end{example}

\begin{example}\label{ex:omega02-from-determinantal}
The leading term of the expansion in Corollary~\ref{cor:BArat} is
\begin{equation}\label{eq:K0-leading}
K_0(p_1,p_2)=[\hbar^0]K(p_1,p_2)=
\frac1{2\Delta_1^{1/2}\Delta_2^{1/2}}\frac{s_1+s_2}{s_1-s_2}\frac{1+t_1t_2}{1-t_1t_2}.
\end{equation}
Then, the determinantal formula gives
\begin{equation}\label{eq:omega02-from-K0}
\omega^{(0)}_2(p_1,p_2)=\Bigl(-K_0(p_1,p_2)\,K_0(p_2,p_1)+K_0(\iota(p_1),p_2)\,K_0(p_2,\iota(p_1))\Bigr)
\frac{d X_1}{X_1}\frac{d X_2}{X_2}.
\end{equation}
This recovers~\eqref{eq:omega02}, after simplification.
\end{example}

\subsection{Possible poles of \texorpdfstring{$K$}{K} and \texorpdfstring{$\omega^{(g)}_n$}{wgn}}
From~\eqref{eq:basicfunctions}--\eqref{eq:R} we get:
\begin{equation}\label{eq:dX}
dX=\frac{\alpha \,\beta\,R(z)}{2\,(\alpha+\beta\,z)(\beta+\alpha\,z)}\frac{dz}{w}.
\end{equation}
It follows that, for generic values of $\alpha,\beta$, that is, for \eqref{eq:nondegeneracy-condition}, the function $X$ has eight critical points corresponding to the four distinct roots of the polynomial $R$. All these critical points contribute to topological recursion and produce poles of the differentials $\omega^{(g)}_n$. It turns out, however, that these are not the only possible poles.

\begin{definition}\label{def:pole-set-P}
For generic position of points $p_1,\dots,p_{n-1}$, we denote by $\cP$ the $(2n+7)$-element set of points on the spectral curve consisting of
\begin{itemize}
\item the eight preimages of the four roots of~$R$ under the elliptic double cover;
\item the point at infinity $(z,w)=(\infty,\infty)$;
\item $n-1$ pairs of preimages of $1/z(p_i)$, $i=1,\dots,n-1$, under the elliptic cover.
\end{itemize}
\end{definition}

\begin{proposition}\label{prop:omega-poles}
For $2g-2+n>0$, the differential $\omega^{(g)}_n$ regarded as a meromorphic $1$-form in the last argument might have poles at $\cP$ only, that is, it is holomorphic in the complement of $\cP$.
\end{proposition}

By determinantal identities, the possible poles of the $\omega$-differentials correspond to possible poles of the $K$-kernel.

\begin{proposition}\label{prop:K-poles}
Possible poles of $K(p_1,p_2)$ lie on the divisors given by $s_1-s_2=0$ and $t_1t_2-1=0$, where $s_i=s(p_i)$ and $t_i=t(p_i)$, and at the points corresponding to the zeros of $R(z)$ and the point at infinity in each of the two arguments.
\end{proposition}

\begin{proof}
Let us examine the proof of Corollary~\ref{cor:BArat} more closely. The denominators in the rational expressions for the coefficients $B_k$ in~\eqref{eq:BArat} can arise from
\begin{itemize}
\item the Taylor expansions of the phases~$\Phi_{X_1}(\sigma_1,\tau_1)$ and $\Phi_{X_2}(\sigma_2,\tau_2)$~\eqref{eq:phase} at the corresponding critical points $(s_1,t_1)$ and $(s_2,t_2)$, respectively;
\item the Taylor expansion of the amplitude~$\cA$~\eqref{eq:amplitude} at the same points;
\item the components of the covariance~\eqref{eq:covariance}.
\end{itemize}
Since $\frac{1}{1-s^2}=\frac{t}{\alpha X}$ and $\frac{1}{1-t^2}=\frac{s}{\beta X}$, it follows that $B_k$ can be written as a polynomial in $(s_1-s_2)^{-1}$, $(t_1t_2-1)^{-1}$, $R(z_1)^{-1}$, $R(z_2)^{-1}$, and also $t_i^{\pm1},s_i^{\pm1},X_i^{\pm1}$. The explicit formulas for $s,t,X$ show that, apart from the possible poles listed in the proposition, individual monomials may have apparent poles at~$z_i=0$, $z_i=\pm1$, and $z_i\in\{-\alpha/\beta,-\beta/\alpha\}$.

The regularity of $K$ at $z_i=0$ follows from the fact that it has an explicit expansion~\eqref{eq:Kregsum} at that point.
Thus the proposition follows from the following lemma.

\begin{lemma}\label{lem:regularity-at-special-points}
If the amplitude~$F$ in the integral~\eqref{eq:2Gintrat} is rational and regular at the points $(s,t)$ whose $z$-coordinates belong to $\{-1,1,-\alpha/\beta,-\beta/\alpha\}$, then the integral is also regular at those points.
\end{lemma}

The apparent poles at $z\in\{-1,1,-\alpha/\beta,-\beta/\alpha\}$ are removed by suitable rescalings of the local coordinates used in the integration.

At the point with~$z=1$ we can set
\begin{equation}
\begin{aligned}
a&=1-s^2=\frac{\alpha X}{t},&b&=1-t^2=\frac{\beta X}{s},
\\
\sigma&=s+\sqrt{\hbar}a\chi,& \tau&=t+\sqrt{\hbar}b\eta
\end{aligned}\label{eq:regularity-coordinates-z-one}
\end{equation}
instead of~\eqref{eq:gaussian-shift-substitution}. Then, for this choice of the local coordinates and for every $r\ge3$, the Taylor coefficients $a^r\arctanh^{(r)}(s)$ and $b^r\arctanh^{(r)}(t)$ of $W_X$ are regular, and the Taylor coefficients of the amplitude $F$ are also regular. The components of the covariance are also regular and its determinant is non-vanishing. Moreover, the Jacobian factor has a double zero at those points that compensates a first order pole of the prefactor $1/X$ in~\eqref{eq:2Gintrat}. So the integral is regular, and, moreover, it has actually a simple zero at the points with $z=\pm1$.

At the point with $z=-\alpha/\beta$, we can choose the local integration coordinates by
\begin{equation}
\sigma=s+\sqrt{\hbar}s\chi,\quad \tau=t+\sqrt{\hbar}\eta.\label{eq:regularity-coordinates-z-minus-alpha-over-beta}
\end{equation}
Then, again, one checks directly that for these coordinates all coefficients in the expansion of the phase and the amplitude, as well as all components of the covariance are regular, and the determinant of the covariance is non-vanishing. This implies the required regularity. The proof for the points with $z=-\beta/\alpha$ is similar, with the exchange of the roles of $s$ and $t$.

This completes the proof of Lemma~\ref{lem:regularity-at-special-points} and also that of Proposition~\ref{prop:K-poles}.
\end{proof}

\begin{proof}[Proof of Proposition~\ref{prop:omega-poles}]
The poles of $K$ at the roots of $R$ and at infinity produce the ``expected'' poles of the $\omega$-differentials. Let us check the contributions of the divisors given by $s_1-s_2=0$ and $t_1t_2-1=0$, respectively. Their images under projection to the plane of the $z$-coordinates are given, respectively, by
\begin{align}
0&=s_1^2-s_2^2=
\frac{\alpha\, (z_1-z_2) (\alpha\, (z_1+ z_2)+\beta\, (z_1 z_2+1))}{(\beta\, z_1+\alpha) (\beta\, z_2+\alpha)},\label{eq:s-difference-divisor}
\\0&=t_1^2t_2^2-1=
\frac{\beta\, (z_1 z_2-1) (\alpha\, (z_1+ z_2)+\beta\, (z_1 z_2+1))}{(\alpha\, z_1+\beta) (\alpha\, z_2+\beta)}.\label{eq:t-product-divisor}
\end{align}
The first one consists of the diagonal $z_1=z_2$ and an extra component given by the equation  $\alpha\, (z_1+ z_2)+\beta\, (z_1 z_2+1)=0$. The second one has the component given by the equation $z_1z_2=1$ and one more component, the same as above.

Notice that $K$ is actually regular along the divisor $\alpha\, (z_1+z_2)+\beta\, (z_1z_2+1)=0$. Indeed, representation~\eqref{eq:fourGaussian} gives the first divisor as the only possible extra divisor, whereas the equivalent twin representation of Remark~\ref{rem:twin-equivalence} gives instead $\beta\, (z_1+z_2)+\alpha\, (z_1z_2+1)=0$. Since these two divisors are distinct for generic $\alpha,\beta$, the kernel is regular along both. Alternatively, the same conclusion follows from a local analysis similar to that in the proof of the lemma above.

The divisor $z_1=z_2$ of $K$ corresponds to potential poles of $\omega^{(g)}_n$ given by $z_i=z_j$ for various pairs $(i,j)$. Its preimage on the Cartesian square of the spectral curve has two components: the diagonal and the $\iota$-antidiagonal. By its definition~\eqref{eq:omega}, $\omega^{(g)}_n$ has a regular power series expansion at the origin, where the two components meet. It follows that $\omega^{(g)}_n$ cannot have poles on either component.

Finally, the divisor $z_1z_2=1$ of $K$ produces the ``expected'' poles of $\omega^{(g)}_n$ given by $z_iz_j=1$. Thus the poles of~$K$ contribute only the poles of $\omega^{(g)}_n$ listed in~$\cP$. This proves Proposition~\ref{prop:omega-poles}.
\end{proof}

\section{Self-duality and the proof of topological recursion}\label{sec:rho-symmetry}

We have already proved the meromorphy of the differentials~$\omega^{(g)}_n$ and provided a list of their possible poles. To complete the formulation of topological recursion, it remains to identify the principal parts at these poles. This is done below in Section~\ref{sec:SymplTR}. The formulation involves symplectic duality, discussed in~\cite{ABDKS05} and reviewed in Appendix~\ref{sec:xy-sympl}, and the symmetry with respect to the involution~$\rho$ introduced below.

\subsection{Involution \texorpdfstring{$\rho$}{rho} and self-duality}
Consider a new involution on the spectral curve
\begin{equation}\label{eq:rhodef}
\rho:(z,w)\mapsto (1/z,w/z^2).
\end{equation}
It commutes with the elliptic involution $\iota$, and for $x=\log X$, $y=\frac{z}{X}$, and $\xi=\frac{z}{w}$, we have:
\begin{equation}\label{eq:rho-action}
\begin{aligned}
s\circ\rho&=\frac1s,\qquad t\circ\rho=\frac1t,\qquad \xi\circ\rho=\xi,
\\y\circ\rho&=-y,\qquad X\circ\rho=-\frac{X}{z^2}=-\frac1{Xy^2},
\\\rho^*\omega^{(0)}_1&=\rho^*(y\,d x)=y\,d (x+\log y^2),
\\\rho_1^*\omega^{(0)}_2&=\rho_2^*\omega^{(0)}_2=\omega^{(0)}_2.
\end{aligned}
\end{equation}
The last two lines of equalities motivate the following theorem serving as the key point in the proof of topological recursion.

\begin{theorem}\label{thm:sympl=rho}
There exists an $\hbar$-deformation $\widetilde\psi$ of the function $\psi(y)=\log(y^2)$ such that the symplectic duals of the differentials $\omega^{(g)}_n$ with respect to the symplectic duality data $(x,y,\widetilde\psi)$ are given by
\begin{equation}\label{eq:rhoomegaidentity}
\omega^{\dagger,(g)}_n=\rho_1^*\dots\rho_n^*\omega^{(g)}_n.
\end{equation}
Explicitly, the required $\hbar$-deformation is given by
\begin{equation}\label{eq:tildepsi-def}
\widetilde\psi(y)=\frac{1}{\cS(\hbar\partial_y)}\log\Bigl(y^2-\frac{\hbar^2}{4}\Bigr)
=\sum_{m\ge0}\frac{B_{2m}}{(2m)!}\hbar^{2m}\partial_y^{2m}\psi
=\log y^2-\sum_{m\ge1}\frac{B_{2m}}{m}\hbar^{2m}y^{-2m}.
\end{equation}
\end{theorem}

By the determinantal identities and according to Definition \ref{def:global-BKP}, the right-hand sides of~\eqref{eq:rhoomegaidentity} form a BKP-integrable system of differentials with Baker--Akhiezer kernel $\rho_1^*\rho_2^*\bK$. The differentials on the left-hand side are also BKP integrable. Thus, to prove the theorem, it suffices to compute the action of the symplectic duality transformation on $\bK$ and compare the result with $\rho_1^*\rho_2^*\bK$. By the integral transform formulas reviewed in Section~\ref{sec:xy-sympl}, the transformation of the kernel is described by integral transformations acting independently on its two arguments, and it therefore suffices to consider the action on only one argument.

Before proceeding with the computations, let us unify the notation for the arguments of the various Gaussian integrals under consideration. Using the shorthand $A$ for $\arctanh$, define
\begin{align}
\Phi_p(\sigma,\tau)&=-\frac{\sigma\tau}{X(p)}+\alpha \,A(\sigma)+\beta\,A(\tau),\label{eq:Phi-p}
\\
\Phi^\vee_p(\sigma,\tau)&=-y(p)\log(\sigma\tau)+\alpha \,A(\sigma)+\beta\,A(\tau),\label{eq:Phi-vee-p}
\end{align}
where $p$ is a point of the spectral curve. The first phase is the same as~\eqref{eq:phase}; the new notation emphasizes its globally meromorphic dependence on the choice of~$p$. Both functions have a critical point at $(\sigma,\tau)=(s(p),t(p))$, and the corresponding critical values are denoted by
\begin{equation}\label{eq:H-critical-values}
H(p)=\Phi_p(s(p),t(p)),\qquad H^\vee(p)=\Phi^\vee_p(s(p),t(p)).
\end{equation}
Straightforward computations then show that the differentials of these functions are
\begin{equation}\label{eq:dH-and-dHvee}
\dd H(p)=y(p)\, \dd x(p),\qquad
\dd H^\vee(p)=-x(p)\, \dd y(p)-\dd\Bigl(y(p)\,\bigl(\log(y(p))-1\bigr)\Bigr).
\end{equation}
In particular, $H=H^{(0)}_1$ is the genus-zero one-point function, up to a constant depending on the choice of branch of $\arctanh$. To make the above definitions precise, we choose a simply connected domain on the spectral curve that avoids the logarithmic singularities of the functions involved, and we fix branches of the logarithms compatible with the above relations. If an analytic identity holds on this domain and both sides extend to globally meromorphic functions, then the identity holds globally. With this convention, we shall not discuss the choices of branches of multivalued functions in the computations below.

We will also need one particular formal power series with rational coefficients defined explicitly by
\begin{multline}\label{eq:Gcal}
\cG(\hbar)=\frac{1}{\sqrt{2\pi\hbar}}\int
 e^{-\theta}\,e^{\frac{1}{\hbar}\bigl(1-\theta-e^{-\theta}\bigr)}
 \dd \theta=\frac{e^{1/\hbar}\hbar^{1/\hbar}\Gamma(1/\hbar)}{\sqrt{2\pi \hbar}}
\\ =\exp\biggl(\sum_{m\ge1}\frac{B_{2m}}{2m\,(2m-1)}\hbar^{2m-1}\biggr)
 =1+\frac{\hbar}{12}+\frac{\hbar^2}{288}-\frac{139\,\hbar^3}{51840}+\dots.
\end{multline}
The expression on the second line implies that this function satisfies $\cG(\hbar)\cG(-\hbar)=1$.

The key computation is contained in the following lemma. For a given rational function $F(\sigma,\tau)$, define
\begin{equation}\label{eq:f-and-fdagger}
\begin{aligned}
f(p)&=\frac{i}{X(p)\,2\pi\hbar}\iint F(\sigma,\tau)\;
e^{\frac1\hbar\bigl(\Phi_{p}(\sigma,\tau)-H(p)\bigr)}\dd\sigma \dd\tau,
\\
f^\dagger(p)&=\frac{1}{X(\rho(p))\,2\pi\hbar}\iint F(1/\sigma,1/\tau)\;
e^{\frac1\hbar\bigl(\Phi_{\rho(p)}(\sigma,\tau)-H(\rho(p))\bigr)}\dd\sigma \dd\tau.
\end{aligned}
\end{equation}

Introduce the integral $\xy$ swap transformation
\begin{equation}\label{eq:Gxy-main}
G_{x,y}f(q)=\frac{1}{\sqrt{2\pi\hbar}}\int f(p)\;
 e^{\frac{1}{\hbar}\left((x(q)-x(p))\,y(q)-\int\limits_p^qy\,d x\right)}\dd x(p)
\end{equation}
reviewed in Appendix~\ref{sec:xy-sympl}.

\begin{lemma}\label{lem:key-identity}
The images of the functions $f$ and $f^{\dagger}$ under the transformations $G_{x,y}$ and $G_{x^\dagger,-y}$, respectively, where $x^\dagger=\log(z^2/X)=x+\log y^2$, are given explicitly by
\begin{align}\label{eq:Gf}
G_{x,y}f(q)&=\cG\bigl(\tfrac{\hbar}{y(q)}\bigr)\frac{i\sqrt{y(q)}}{2\pi\hbar}\iint \frac{F(\sigma,\tau)}{\sigma\tau}
e^{\frac1\hbar\bigl(\Phi_q^\vee-H^\vee(q)\bigr)}\dd \sigma\dd\tau,
\\\label{eq:Gfdag}
G_{x^\dagger,-y}f^\dagger(q)&=\cG\bigl(-\tfrac{\hbar}{y(q)}\bigr)\frac{i\sqrt{y(q)}}{2\pi\hbar}\iint \frac{F(\sigma,\tau)}{\sigma\tau}
e^{\frac1\hbar\bigl(\Phi_q^\vee-H^\vee(q)\bigr)}\dd \sigma\dd\tau.
\end{align}
As a corollary, these functions differ by a universal factor independent of a choice of~$F$:
\begin{equation}\label{eq:Gcal-ratio}
\frac{G_{x,y}f(q)}{G_{x^\dagger,-y}f^\dagger(q)}
=\frac{\cG\bigl(\tfrac{\hbar}{y(q)}\bigr)}{\cG\bigl(-\tfrac{\hbar}{y(q)}\bigr)}
=\cG\bigl(\tfrac{\hbar}{y(q)}\bigr)^2.
\end{equation}
\end{lemma}

Note that the factor in the last assertion can be represented in the form
\begin{equation}\label{eq:G-vs-tildePsi}
\cG\bigl(\tfrac{\hbar}{y}\bigr)^2=\exp\biggl(2\sum_{m\ge1}\frac{B_{2m}}{2m\,(2m-1)}\hbar^{2m-1}y^{-2m+1}\biggr)
=\exp\biggl(\frac1\hbar\int^y(\widetilde\psi(\theta)-\psi(\theta))d\theta\biggr),
\end{equation}
where $\widetilde\psi$ is exactly the series in~\eqref{eq:tildepsi-def}.

\begin{proof}
Combining the integral representation of $f$ with the integral transformation $G_{x,y}$ and taking into account that $X=e^x$ and $\int_p^q y\,dx=H(q)-H(p)$, we write the result as a triple Gaussian integral:
\begin{equation}\label{eq:triple-gaussian}
\begin{aligned}
G_{x,y}f(q)&
=\frac{i}{(2\pi\hbar)^{\frac32}}\iiint \frac{F(\sigma,\tau)}{X(p)}\,
e^{\frac{1}{\hbar}\bigl(-\frac{\sigma\tau}{X(p)}+\alpha \,A(\sigma)+\beta\,A(\tau)-H(p)\bigr)} \dd\sigma \dd\tau
\\&\hskip3cm\times
 e^{\frac{1}{\hbar}\bigl((x(q)-x(p))\,y(q)+\int\limits_q^py\,d x\bigr)}\dd x(p)
\\&
=\frac{i}{(2\pi\hbar)^{\frac32}}\iiint \frac{F(\sigma,\tau)}{e^x}\,
e^{\frac{1}{\hbar}\bigl(-\sigma\tau e^{-x}+\alpha \,A(\sigma)+\beta\,A(\tau)-H(q)
+(x(q)-x)\,y(q)\bigr)}\dd\sigma \dd\tau\dd x
\\&=\frac{i}{(2\pi\hbar)^{\frac32}}\iiint
F(\sigma,\tau)\,e^{\frac{1}{\hbar}\bigl(
-y(q)\,\log(\sigma\tau)+\alpha \,A(\sigma)+\beta\,A(\tau)-H^\vee(q)\bigr)}
 \dd\sigma \dd\tau
 \\&\hskip3cm\times
 e^{-x}\,e^{\frac{1}{\hbar}\bigl(-\sigma\tau e^{-x}-x\,y(q)+y(q)\,(\log\frac{\sigma\tau}{y(q)}+1)\bigr)}
 \dd x.
\end{aligned}
\end{equation}
We distributed the constant terms between the two phases such that they both take zero values at the corresponding critical points. Using Fubini factorization, let us compute the integral over~$x$ first. The critical point of the phase is given by the equation
\begin{equation}\label{eq:x-critical-equation}
\partial_x\bigl(-\sigma\tau e^{-x}-y(q)\,x\bigr)=0
\end{equation}
which implies $x_{\rm crit}=\log\frac{\tau\sigma}{y(q)}$. The change of variables $x=\log\frac{\tau\sigma}{y(q)}+\theta$ shifts the critical point to the origin and the integral reduces to
\begin{equation}\label{eq:x-integral-gives-Gcal}
\begin{aligned}
\frac{1}{\sqrt{2\pi\hbar}}\int
 e^{-x}&\,e^{\frac{1}{\hbar}\bigl(-\sigma\tau e^{-x}+y(q)\,(-x+\log\frac{\sigma\tau}{y(q)}+1)\bigr)}
 \dd x
 \\&=
\frac{y(q)}{\sigma\tau}\frac{1}{\sqrt{2\pi\hbar}}\int
 e^{-\theta}\,e^{\frac{y(q)}{\hbar}\bigl(1-\theta-e^{-\theta}\bigr)}
\dd \theta=\frac{\sqrt{y(q)}}{\sigma\tau}\cG\bigl(\tfrac{\hbar}{y(q)}\bigr).
\end{aligned}
\end{equation}
 We get finally
 \begin{equation}\label{eq:Gfproof}
 G_{x,y}f(q)=\cG\bigl(\tfrac{\hbar}{y(q)}\bigr)\;
 \frac{i\sqrt{y(q)}}{2\pi\hbar}\iint
\frac{F(\sigma,\tau)}{\sigma\tau}\,e^{\frac{1}{\hbar}\bigl(
-y(q)\,\log(\sigma\tau)+\alpha \,A(\sigma)+\beta\,A(\tau)-H^\vee(q)\bigr)}
 \dd\sigma \dd\tau
 \end{equation}
which coincides with~\eqref{eq:Gf}.

For~\eqref{eq:Gfdag}, the same computations, with $y$ replaced by $-y$, lead to
\begin{equation}\label{eq:Gfdag-intermediate}
 G_{x^\dagger,-y}f^\dagger(q)=\cG\bigl(-\tfrac{\hbar}{y(q)}\bigr)\;
 \frac{\sqrt{-y(q)}}{2\pi\hbar}\iint
\frac{F(\tfrac{1}{\sigma},\tfrac{1}{\tau})}{\sigma\tau}\,e^{\frac{1}{\hbar}\bigl(
y(q)\,\log(\sigma\tau)+\alpha \,A(\sigma)+\beta\,A(\tau)-H^\vee(\rho(q))\bigr)}
 \dd\sigma \dd\tau
\end{equation}
The phase in this integral differs from that in~\eqref{eq:Gfproof}, up to a constant, only in the sign of the term containing $y(q)$. To correct this sign, we apply the change of variables $\sigma\mapsto1/\sigma$, $\tau\mapsto1/\tau$. Using $d\sigma/\sigma=-d(\sigma^{-1})/\sigma^{-1}$ and $d\tau/\tau=-d(\tau^{-1})/\tau^{-1}$, together with the fact that $\arctanh(\sigma)$ and $\arctanh(\sigma^{-1})$ differ by a constant, we finally obtain~\eqref{eq:Gfdag}.

\end{proof}

\begin{proof}[Proof of Theorem~\ref{thm:sympl=rho}]
Taking~\eqref{eq:G-vs-tildePsi} into account, Lemma~\ref{lem:key-identity} can be rephrased as saying that $f^\dag \sqrt{dx^\dag}$ is the result of applying the symplectic integral transformation $\bG_{x,y,\widetilde\psi}$ of Appendix~\ref{sec:xy-sympl} to~$f\sqrt{dx}$:
\begin{equation}\label{eq:fdag-is-sympl-transform}
f^\dagger\sqrt{dx^\dag}=\bG_{x,y,\widetilde\psi}\;f\sqrt{dx}.
\end{equation}
On the other hand, since $dx^\dag=-\rho^*dx$, we have $\rho^*(f^\dagger\sqrt{dx^\dag})=i (\rho^*f^\dagger)\sqrt{dx}$, and $i\,\rho^*f^\dagger$ is represented by an integral similar to that for~$f$, but with an amplitude whose arguments~$\sigma$ and~$\tau$ are inverted. Let us apply this observation to the action of the symplectic integral transformation on the kernel~$\bK$. The kernel is given by a quadruple Gaussian integral with amplitude $\cA(\sigma_1,\tau_1,\sigma_2,\tau_2)$. By the lemma, the half-differential
\begin{equation}\label{eq:rho-G-kernel}
\rho_1^*\bG^{(p_1)}_{x,y,\widetilde\psi}\bK(p_1,p_2)
\end{equation}
is also represented by a similar quadruple Gaussian integral, but with the amplitude
\begin{equation}\label{eq:amplitude-inversion-is-twin}
\cA\bigl(\tfrac{1}{\sigma_1},\tfrac{1}{\tau_1},\sigma_2,\tau_2\bigr)=
\frac{\frac{1}{\sigma_1}+\sigma_2}{\frac{1}{\sigma_1}-\sigma_2}
\frac{1+\frac{\tau_2}{\tau_1}}{1-\frac{\tau_2}{\tau_1}}
=\cA^{\rm tw}(\sigma_1,\tau_1,\sigma_2,\tau_2).
\end{equation}
By Remark~\ref{rem:twin-equivalence}, this integral still gives the original~$\bK$. Therefore we have proved
\begin{equation}\label{eq:G-kernel-rho1}
\bG^{(p_1)}_{x,y,\widetilde\psi}\bK(p_1,p_2)=\rho_1^*\bK(p_1,p_2),
\end{equation}
and, similarly,
\begin{gather}
\bG^{(p_2)}_{-x,y,-\widetilde\psi}\bK(p_1,p_2)=\rho_2^*\bK(p_1,p_2),\label{eq:G-kernel-rho2}
\\
\bG^{(p_1)}_{x,y,\widetilde\psi}\bG^{(p_2)}_{-x,y,-\widetilde\psi}\bK(p_1,p_2)=\rho_1^*\rho_2^*\bK(p_1,p_2),\label{eq:G-kernel-rho12}
\end{gather}
Applying the determinantal identities, we obtain the desired formula in the theorem.
\end{proof}

In fact, the arguments in the proof of Theorem~\ref{thm:sympl=rho} prove even more. Namely, we have

\begin{theorem}\label{thm:sympl=rho-mixed}
The mixed intermediate $(k,n-k)$ differentials of symplectic duality with data $(x,y,\widetilde\psi)$ as in Theorem~\ref{thm:sympl=rho} are given by
\begin{equation}\label{eq:mixed-differentials}
\omega^{(g)}_{k,n-k}=\rho_{k+1}^*\dots\rho_n^*\omega^{(g)}_n,
\end{equation}
with the convention $\omega^{(g)}_{n,0}=\omega^{(g)}_{n}$, $\omega^{(g)}_{0,n}=\omega^{\dagger,(g)}_{n}$. Moreover, they satisfy determinantal identities: set
\begin{equation}\label{eq:hat-p}
\widehat p_i=\left\{
\begin{array}[c]{ll}
p_i,&1\le i\le k,\\
\rho(p_i),&k<i\le n.
\end{array}\right.
\end{equation}
Then,\
\begin{equation}\label{eq:detid-mixed}
\omega_{k,n-k}(p_1,\dots,p_n)=
{\rm Alt}_1\cdots{\rm Alt}_{n-1}(-1)^{n-1}
\sum_{\sigma\in\operatorname{cycl}(n)}\prod_{i=1}^{n}\bK(\widehat p_i,\widehat p_{\sigma(i)}),
\quad n\ge2,
\end{equation}
where ${\rm Alt}_i={\rm Id}-\iota^*$. (The case $n=1$ is covered by Theorem~\ref{thm:sympl=rho}.)
\end{theorem}

\subsection{Symplectic topological recursion}\label{sec:SymplTR}

We are now ready to give a precise formulation of topological recursion for the differentials~$\omega^{(g)}_n$ of the present paper.
Let us make a simple but important observation: for generic fixed values of the remaining arguments, the set supporting the $2n+7$ possible poles of stable $\omega^{(g)}_n$ in any one argument is disjoint from its image under the involution $\rho$. Consequently, the mixed $(n-1,1)$ differentials
\begin{equation}\label{eq:on1rho}
\omega^{(g)}_{n-1,1}=\rho_n^*\omega^{(g)}_n
\end{equation}
from Theorem~\ref{thm:sympl=rho-mixed} are holomorphic at the poles of~$\omega^{(g)}_n$ in the last argument.

\begin{definition}\label{def:sympl-loop-equations} We say that the collection of differentials $\omega^{(g)}_n$ satisfies the \emph{loop equations of symplectic topological recursion} for the data $(x,y,\widetilde \psi)$ at a point if the mixed $(n-1,1)$ differentials~$\omega^{(g)}_{n-1,1}$ of symplectic duality are holomorphic at that point.
\end{definition}

We conclude:

\begin{theorem}\label{thm:loop-equations-hold}
For $2g-2+n>0$, the differentials $\omega^{(g)}_n$ for the spin enumeration of hypermaps satisfy the symplectic topological recursion loop equations with $x=\log X$, $y=z/X$, and $\widetilde \psi$ as in~\eqref{eq:tildepsi-def} at every singular point in the last argument.
\end{theorem}

Let us clarify the meaning of this statement. By the definition of symplectic duality in Appendix~\ref{sec:xy-sympl}, the mixed differentials can be given by a combinatorial expression of the form
\begin{equation}\label{eq:mixed-as-differential-polynomial}
\omega^{(g)}_{n-1,1}(p_{\set{n-1}},p)=\sum_{j\ge0}\bigl(-d\tfrac{1}{dx^\dagger}\bigr)^j[v^j]\cW^{(g)}_{n}(v,p),
\end{equation}
where $\cW^{(g)}_{n}(v,p)$ is expressed in terms of differentials $\omega^{(g')}_{n'}$ for various $(g',n')$ and depends polynomially on $v$ --- its explicit formula is given in Equation~\eqref{eq:curly-W-definition}. The mixed differentials are discussed in detail in Section~\ref{sec:mixed-differentials}, here it is merely a notation for the expression on the right hand side of~\eqref{eq:mixed-as-differential-polynomial}.

Moreover, the summand with $j=0$ in~\eqref{eq:mixed-as-differential-polynomial} is exactly $\omega^{(g)}_n$, whereas the summands with $j>0$ involve differentials $\omega^{(g')}_{n'}$ with $2g'-2+n'<2g-2+n$, which are regarded as having been computed in previous steps of the recursion. Therefore, define
\begin{equation}\label{eq:tildeomega-def}
\widetilde \omega^{(g)}_n=-\sum_{j\ge1}\bigl(-d\tfrac{1}{dx^\dagger}\bigr)^j[v^j]\cW^{(g)}_{n}(v,p).
\end{equation}
These differentials are known from the previous steps of the recursion, and we have
\begin{equation}\label{eq:omega-split}
\omega^{(g)}_n|_{p_n = p}=\widetilde\omega^{(g)}_n+\omega^{(g)}_{n-1,1}
\end{equation}
Thus $\omega^{(g)}_n$ and $\widetilde\omega^{(g)}_n$ have the same principal parts at all $2n+7$ singular points of $\omega^{(g)}_n$ in its last argument.

In this way, the loop equations determine the principal parts at all poles of $\omega^{(g)}_n$ and thereby determine the differential itself uniquely up to a holomorphic summand. This summand is fixed by the following property.

\begin{proposition}\label{prop:vanishing-residues-periods}
For $2g-2+n>0$, the differentials $\omega^{(g)}_n$ have vanishing residues and periods with respect to all variables. Consequently, they are total differentials of meromorphic functions.
\end{proposition}

\begin{proof}
First, the differentials $\widetilde\omega^{(g)}_n$ are manifestly total differentials by the form of~\eqref{eq:tildeomega-def}: every summand with $j\ge1$ is a total differential. Since $\omega^{(g)}_n$ has the same principal parts at its poles as~$\widetilde\omega^{(g)}_n$, it has the same residues, namely zero. Thus, the vanishing of residues is a general consequence of the form of the recursion.

By contrast, the vanishing of periods does not follow from the general properties of symplectic topological recursion but is specific to the problem under consideration. By~\eqref{eq:on1rho}, the combination
\begin{equation}\label{eq:omega-minus-rho-omega}
\omega^{(g)}_n-\rho_n^*\omega^{(g)}_n=\omega^{(g)}_n-\omega^{(g)}_{n-1,1}=\widetilde\omega^{(g)}_n
\end{equation}
has vanishing periods in the last argument. On the other hand, it follows directly from~\eqref{eq:rhodef} that $\rho^*(dz/w)=-dz/w$. Thus $\rho^*$ acts on the first cohomology of the curve by multiplication by~$-1$, and its action on the first homology is also multiplication by~$-1$. If $P$ is a period of $\omega^{(g)}_n$, then the corresponding period of $\rho_n^*\omega^{(g)}_n$ is $-P$. Hence the corresponding period of $\omega^{(g)}_n-\rho_n^*\omega^{(g)}_n$ is $2P$. Since the latter differential has vanishing periods, $P=0$.
\end{proof}

The procedure for recovering the differentials from their principal parts can be written in terms of the standard projection operator. Let $B$ be a Bergman kernel, that is, a symmetric bidifferential with a pole on the diagonal of biresidue~$1$ and no other poles. Without a period normalization, it is defined up to a global holomorphic bidifferential proportional to~$\frac{dz_1}{w_1}\frac{dz_2}{w_2}$. An example is
\begin{equation}\label{eq:B}
B=\frac{(\xi_1+\xi_2)^2}{4\xi_1\xi_2}\;\frac{dz_1 dz_2}{(z_1-z_2)^2},
\qquad \xi_i=\frac{z_i}{w_i}.
\end{equation}

In the traditional setting of topological recursion, the Bergman kernel is fixed by requiring its $\fA$-periods to vanish for a chosen basis of $\fA$- and $\fB$-cycles on the curve. In our case, however, this normalization is unnecessary, as the following proposition states.

\begin{proposition}\label{prop:projection-property}
For $2g-2+n>0$, for any choice of the Bergman kernel the differentials $\omega^{(g)}_n$ obey the projection property
\begin{align}
\omega^{(g)}_n(p_{\set{n}})
  &=\sum_{q\in\cP}\res_{p=q}\omega^{(g)}_n(p_{\set{n-1}},p)\int^pB(\cdot,p_n)
\label{eq:proj1}
\\&=\sum_{q\in\cP}\res_{p=q}\widetilde\omega^{(g)}_n(p_{\set{n-1}},p)\int^pB(\cdot,p_n)
\label{eq:proj2}
\end{align}
where $\cP$ is the collection of all $2n+7$ possible poles.
\end{proposition}

\begin{proof}
The second equality follows because the residues depend only on the polar parts of the differentials, which are the same for the two differentials. To prove the first equality, denote by $\Pi$ the $4$-gon obtained by cutting the surface along chosen $\fA$- and $\fB$-cycles. For $p\in\Pi$, let $S_p(\cdot)$ be a primitive of $B(\cdot,p)$ in $\Pi$. Applying the residue theorem in $\Pi$ to the differential $\omega^{(g)}_n(p_{\set{n-1}},\cdot)S_{p_n}(\cdot)$, we obtain
\begin{multline}\label{eq:residue-theorem-on-4gon}
\omega^{(g)}_n(p_{\set{n}})=\sum_{q\in\cP}\res_{p=q}\omega^{(g)}_n(p_{\set{n-1}},p)S_{p_n}(p)\\
+\frac{1}{2\pi i}\int\limits_{\fA}\!\!B(\cdot,p_n)\;
\int\limits_{\fB}\!\omega^{(g)}_n(p_{\set{n-1}},\cdot)
-\frac{1}{2\pi i}\int\limits_{\fB}\!\!B(\cdot,p_n)\;
\int\limits_{\fA}\!\omega^{(g)}_n(p_{\set{n-1}},\cdot).
\end{multline}
Since the periods of $\omega^{(g)}_n$ vanish, the terms on the second line also vanish, and~\eqref{eq:proj1} follows.
\end{proof}

In conclusion, \emph{a compact form of the topological recursion relation for the differentials $\omega^{(g)}_n$ consists of the explicit expression~\eqref{eq:tildeomega-def} for~$\widetilde\omega^{(g)}_n$ together with the projection formula~\eqref{eq:proj2}}.

\begin{remark}\label{rem:comparison-with-other-TR}
It is worth comparing the loop equations of symplectic topological recursion with other versions of (degenerate) topological recursion, which will be done in~\cite{ABDKS-SymplTR}. One can show that, at the eight nondegenerate critical points of $X$, these loop equations are equivalent to the usual CEO linear and quadratic loop equations~\cite{BEO,BS}. At infinity, which is a degenerate ramification point, they are equivalent to the loop equations of Bouchard--Eynard~\cite{BE-2} (but not GenTR~\cite{ABDKS07}) topological recursion. Finally, at the moving antidiagonals $z_iz_j=1$, we have no alternative formulation of the loop equations other than the one using symplectic duality.
\end{remark}

A final remark on topological recursion concerns the $\iota^*$-symmetry. All differentials $\omega^{(g)}_n$ and also $\widetilde \omega^{(g)}_n$ are odd with respect to the elliptic involution $\iota$. It follows that the Bergman kernel in the projection formula~\eqref{eq:proj2} can be replaced by its normalized antisymmetrization (with respect to either argument), without changing the result of the projection. For example, the normalized antisymmetrization of the differential~\eqref{eq:B} is
\begin{equation}\label{eq:Btilde}
\widetilde B=\frac{\xi_1^2+\xi_2^2}{4\xi_1\xi_2}\frac{dz_1dz_2}{(z_1-z_2)^2}.
\end{equation}
The advantage of using $\widetilde B$ instead of $B$ is that the contributions of $\iota$-symmetric points to the residue expression are equal. It is therefore sufficient to compute only one of them and multiply it by~$2$. This halves the number of residues that must be computed, with the exception of infinity, which is a fixed point of the involution.

\appendix

\section{The sign of an odd hypermap}\label{appx:spin-sign}

In the spin enumeration of hypermaps, we assign a weight $w(\gamma)$ taking values in $\{0,1,-1\}$ to each hypermap $\gamma$, such that $w(\gamma)\ne0$ if and only if the degrees of all faces and vertices of the graph are \emph{odd}. The definition of the weight is implicit in~\cite{EOP,Gun}. Here we provide an independent explicit definition.

Recall that a hypermap can be encoded by a triple of permutations $\sigma,\alpha,\phi$ satisfying $\sigma\alpha\phi=\1$, where $\1$ is the identity permutation. The cycle types of these permutations are the collections of degrees of white vertices, black vertices, and faces, respectively. The definition of the weight is formulated in terms of the combinatorics of the group algebra of the symmetric group and a certain extension of it.

The \emph{Clifford algebra} $\Lambda_d$ is the unital associative algebra over~$\C$ generated by $\xi_1,\dots,\xi_d$, subject to the relations
\begin{equation}	
\xi_i^2=1,\qquad\text{and}\quad\xi_i\xi_j=-\xi_j\xi_i\text{~for~}i\ne j.
\label{eq:clifford-relations}
\end{equation}
The monomials $\xi_1^{a_1}\dots\xi_d^{a_d}$, $a_i\in\{0,1\}$, form a $\C$-basis of~$\Lambda_d$.

The \emph{Sergeev algebra}~\cite{Sergeev} is the semidirect product $\cY_d=\Lambda_d\rtimes\C S(d)$ determined by the commutation relations
\begin{equation}	
s\,\xi_{i}=\xi_{s(i)}\,s,\quad s\in S(d),~ \xi_i\in\Lambda_d.
\label{eq:sergeev-relations}
\end{equation}
A $\C$-basis of $\cY_d$ is formed by monomials of the form $\xi_1^{a_1}\dots\xi_d^{a_d}\,s$, $a_i\in\{0,1\}$, $s\in S(d)$.

For $s\in S(d)$, we define the \emph{normalized Clifford conjugacy average} $\nu(s)\in\cY_d$ by
\begin{equation}	
\nu(s)=\frac{1}{2^{\frac{d+\ell(s)}{2}}}\sum_{J=\{j_1,\dots,j_\ell\}\subset\{1,\dots,d\}}
\xi_{j_1}\dots\xi_{j_\ell}s\; \xi_{j_{\ell}}\dots\xi_{j_1}
\label{eq:nazarov-element}
\end{equation}
where $\ell(s)$ is the number of cycles in~$s$.

For a collection of permutations $s_1,\dots,s_k$ with all cycles of odd lengths in each of them and satisfying $s_1\dots s_k=\1$, we define the weight $w(s_1,\dots,s_k)$ of this collection by the equality
\begin{equation}
	w(s_1,\dots,s_k)\coloneqq \nu(s_1)\dots \nu(s_k).
\label{eq:weight-of-collection}
\end{equation}
The condition $s_1\dots s_k=\1$ implies that $w$ takes values in $\Lambda_d$ (that is, in $\Lambda_d\1\subset\cY_d$ where $\1\in S(d)$ is the identity permutation).

\begin{lemma}\label{lem:weight-is-pm-one} We have, in fact, $w(s_1,\dots,s_k)=\pm 1 \cdot \1$.
\end{lemma}

In particular, in the spin enumeration of hypermaps the definition of the weight is used with $k=3$.

\begin{proof}
	First, note that
	\begin{equation}\label{eq:xinucom}
		\forall i,\,s \qquad \xi_i\nu(s)=\nu(s)\xi_i.
	\end{equation}
	This is the same as stating that $\nu(s)$ is invariant under conjugation by $\xi_i$. Direct computation shows that every $J$-summand with $J$ not containing $i$ in the RHS of~\eqref{eq:nazarov-element} is mapped under such conjugation exactly to the $(J\cup \{i\})$-summand and vice versa; thus this is a bijection on the set of summands.
		
	Second, let us see that
	\begin{equation}		\label{eq:wl1}
		w=\lambda \1, \qquad\lambda\in\mathbb{C},	
	\end{equation}		
	i.e. as a Clifford algebra element (w.r.t. the identification of $\Lambda_d$ with $\Lambda_d\1\subset\cY_d$) $w$ is a scalar (a grade-zero element). Note that $w$ is obviously even with respect to Clifford algebra grading, and that~\eqref{eq:xinucom} immediately implies that conjugation by any Clifford algebra generator $\xi_i$ leaves $w$ invariant. Since $w$ is even and central, it is easy to see that it must be a scalar.
	
	Consider $\C$-linear anti-involution of $\cY_d$ given by $\xi_i^*=\xi_i$, $s^*=s^{-1}$. Let us see that
	\begin{equation}		\label{eq:wws}
	w w^*=\1.
	\end{equation}
	In order to see this, let us prove that for any $s$ consisting only of cycles of odd length, we have
	\begin{equation}\label{eq:nunu1}		
	\nu(s)\nu(s^{-1})=\1.
	\end{equation}
	Note that~\eqref{eq:wl1} for $k=2$ implies that $ \nu(s)\nu(s^{-1})$ is a scalar in terms of the Clifford algebra, and thus
	\begin{equation}		
	\nu(s)\nu(s^{-1}) = [\1] \nu(s)\nu(s^{-1}).\label{eq:nunu-is-its-scalar-part}
	\end{equation}
	On the other hand,
	\begin{equation}\label{eq:nuss1}
		[\1] \nu(s) s^{-1} = 2^{-\frac{\ell(s)+d}{2}}\sum_{J=\{j_1,\dots,j_\ell\}\subset\{1,\dots,d\}} [\1]	\xi_{j_1}\dots\xi_{j_\ell}s\; \xi_{j_{\ell}}\dots\xi_{j_1} s^{-1} = 2^{\frac{\ell(s)-d}{2}},
	\end{equation}
	since a term in the sum above contributes to the scalar part if and only if $s(J)=J$, and moreover it contributes exactly $1$ due to $s|_J$ being an even permutation due to $s$ consisting only of cycles of odd length; thus the whole sum is just equal to $2^{\ell(s)}$, since $J$ can either contain the whole support of some cycle of $s$ or not. Now
	\begin{align}
		[\1] \nu(s)\nu(s^{-1}) &= 2^{-\frac{\ell(s)+d}{2}}\sum_{J}[\1] \nu(s)\xi_J s^{-1} \xi_J^{-1} \mathop{=}^{\eqref{eq:xinucom}} 2^{-\frac{\ell(s)+d}{2}}\sum_{J}[\1] \xi_J \nu(s) s^{-1} \xi_J^{-1}\label{eq:scalar-part-of-nunu-by-conjugation} \\
		&= 2^{d-\frac{\ell(s)+d}{2}}[\1] \nu(s) s^{-1}  \mathop{=}^{\eqref{eq:nuss1}} 1.\label{eq:scalar-part-of-nunu-equals-one}
	\end{align}
	Here $\xi_J = \xi_{j_1}\dots\xi_{j_\ell}$ and we assume that $j_1<\dots<j_\ell$. To pass to the second line we used the fact that the scalar part is invariant w.r.t. the conjugation by any $\xi_i$ and thus by any $\xi_J$, and the fact that there are $2^d$ subsets $J$. Thus we have~\eqref{eq:nunu1}. Then \eqref{eq:wws} immediately follows due to $\nu(s)^*=\nu(s^{-1})$, which is clear from the definitions of $\nu$ and of the $*$-operator, and due to the $*$-operator being an anti-involution.

	From \eqref{eq:wl1} we get $w^*=w$, and this together with ~\eqref{eq:wws} gives $w^2=\1$;  recalling~\eqref{eq:wl1} again, this implies the statement of the lemma.

\end{proof}

\begin{remark}
	Note that in~\eqref{eq:spinNumbers} we treat $w$ as $\pm 1 \in \mathbb{C}$, i.e. as its coefficient in front of $\1$.
\end{remark}

\section{\texorpdfstring{$\xy$}{x-y} swap and symplectic duality}\label{sec:xy-sympl}

In this appendix, we review the $\xy$ swap and symplectic duality relations. The $\xy$ swap relation was originally  conjectured in~\cite{borot2023functionalrelationshigherorderfree}; the simplified formula that we use was derived in~\cite{Hock-formula,ABDKS01}; a proof in genus $0$ was given in~\cite{Hock-genus0} and in full in~\cite{ABDKS01}. We heavily use its links to integrability developed in~\cite{ABDKS03}. Symplectic duality was originally developed in~\cite{BDKS04}, and then with the introduction of LogTR~\cite{ABDKS04} inspired by~\cite{Hock-quantum-diLog}, fully revisited in~\cite{ABDKS05}.

Let $\{\omega^{(g)}_n\}$ be a collection of symmetric meromorphic differentials on Cartesian powers of the spectral curve, all regular on the diagonals except for $\omega^{(0)}_2$, which has a pole on the diagonal with biresidue~$1$. The $\xy$ swap and symplectic duality transformations produce new systems of differentials denoted by $\{\omega^{\vee,(g)}_n\}$ and $\{\omega^{\dagger,(g)}_n\}$, respectively. The transformations themselves are defined independently of the origin of the differentials, topological recursion, or KP/BKP integrability. Since the literature contains some discrepancies in sign conventions, we fix the conventions used below.

\begin{itemize}
\item The data defining the $\xy$ swap transformation consist of two functions $x$ and $y$ on the spectral curve such that $dx$ and $dy$ are meromorphic differentials (thus, $x$ and $y$ may have logarithmic singularities).

The data defining the symplectic transformation consist of functions $x$ and $y$ such that $dx$ and $y$ are meromorphic, together with a function $\psi(\theta)$ and its $\hbar^2$-deformation, that is, a formal series $\widetilde\psi(\theta,\hbar)=\psi(\theta)+\sum_{k\ge1}\hbar^{2k}\psi_k(\theta)$ such that $\psi'(\theta)$ and each $\psi_k(\theta)$, $k\ge1$, are rational.

\item We do not include the $(0,1)$ differentials in the systems to which these transformations are applied. Thus, the systems under consideration are
 $\{\omega^{(g)}_n\}_{\substack{g\ge0,~n\ge1\\(g,n)\ne(0,1)}}$, $\{\omega^{\vee,(g)}_n\}_{\substack{g\ge0,~n\ge1\\(g,n)\ne(0,1)}}$, and $\{\omega^{\dagger,(g)}_n\}_{\substack{g\ge0,~n\ge1\\(g,n)\ne(0,1)}}$. The $(0,1)$ differentials do not participate in the relations by themselves, but only through explicit contributions from the functions $x$ and $y$. Accordingly, we set
\begin{align}\label{eq:omega-n}
  \omega_n=\sum_{g\ge0,\;(g,n)\ne(0,1)}\hbar^{2g-2+n}\,\omega^{(g)}_n ,
\end{align}
and similarly for $\omega^\vee_n$, $\omega^\dagger_n$, and the corresponding extended $\Omega$-differentials. Notice, however, that in the setting of topological recursion, our choice of signs corresponds to
\begin{equation}\label{om01signs}
\omega^{(0)}_{1}=y\,dx,
\qquad \omega^{\vee,(0)}_1=y^\vee\,dx^\vee=-x\,dy,
\qquad \omega^{\dagger,(0)}_1=y^\dagger\,dx^\dagger=y\,d(x+\psi(y)).
\end{equation}
\end{itemize}
Note that the topological expansion as in Equation~\eqref{eq:omega-n} is not a strictly necessary assumption. One can define both transformations for a non-topological expansion perturbative in $\hbar$,   $\omega_n=\sum_{h\ge0}\hbar^{h}\,\omega^{\left<h\right>}_n$.

\begin{remark}\label{rem:Z2-equivariant-context}
	A particular goal of this appendix is to discuss $\xy$ swap and symplectic duality in the presence of a holomorphic involution. The natural context for this setup is  $\mathbb{Z}_2$-equivariant topological recursion developed in~\cite{GKL}. However, we do not review this approach, for the following reason.
	One of the main messages of this appendix is that the definition of $\mathbb{Z}_2$-equivariant topological recursion in the case of an odd Bergman kernel is to a large extent obsolete in the sense that all algebraic tools coming with the theory of $\xy$ swap and symplectic duality allow us to deal with it as a usual topological recursion, with a caveat that the roles of the bi-differential used in the projection formula and initial $\omega^{(0)}_{2}$ used in the loop equations are quite different and they should not necessarily coincide.

	Another message that we want to communicate here is that the natural integrability context in this case is provided by BKP integrability~\cite{DJKM} --- the $\mathbb{Z}_2$-equivariant topological recursion with an odd Bergman kernel produces, under certain assumptions, the BKP tau functions, and  related BKP determinantal identities are preserved by the involved transformations.
\end{remark}

\subsection{Extended differentials}\label{sec:extended-differentials}

The extended differentials package the graph combinatorics of the $\xy$ swap and symplectic dualities.

\begin{definition}[{\cite[Def.~3.2]{ABDKS05}}]\label{def:extended-differentials}
The \emph{extended
$n$-point half-differential} is
\begin{align}\label{eq:Omega-graph}
  \Omega_n(p^+_{\set n},p^-_{\set n})
  =\prod_{i=1}^n\frac{\sqrt{dz^+_idz^-_i}}{z^+_i-z^-_i}
   \sum_{\Gamma\ \mathrm{connected}}\frac{1}{|\Aut\Gamma|}
   \prod_{e\in E(\Gamma)}\ \int_{p^-_{e(1)}}^{p^+_{e(1)}}\!\!\!\!\cdots
   \int_{p^-_{e(|e|)}}^{p^+_{e(|e|)}}\!\!\!\widetilde\omega_{|e|},
\end{align}
a sum over connected graphs with $n$ labelled vertices and multiedges of
arbitrary index $|e|\ge1$, the $j$-th leg of the multiedge $e$ being attached to
the vertex $e(j)$. Each multiedge carries the differential \eqref{eq:omega-n}
of its index, integrated from $p^-$ to $p^+$ in each of its legs, with
$\widetilde\omega_{|e|}=\omega_{|e|}$ except that
\begin{align}\label{eq:regularised-om02}
  \widetilde\omega^{(0)}_2\coloneq \omega^{(0)}_2-\frac{dz_1\,dz_2}{(z_1-z_2)^2},
  \qquad\text{when the two legs of a multiedge sit at one vertex.}
\end{align}
Finally, $z$ is any nonconstant meromorphic function on~$\Sigma$. The $\Omega$-differentials are independent of this choice, although choosing such a function is necessary to state the definition.

If the $\omega$-differentials have nontrivial periods or residues, then, to avoid ambiguity in the choice of branches of the integrals, we assume that $p^+_i$ and $p^-_i$ are close to the diagonal. Also, the half-differentials $\sqrt{dz_i^+dz_i^-}$ in~\eqref{eq:Omega-graph} carry one
sign ambiguity per variable. We fix it once and for all by
\begin{align}\label{eq:sqrt-normalisation}
  \sqrt{dz^+dz^-}\longrightarrow dz\qquad\text{as }p^\pm\to p .
\end{align}
\end{definition}

Notice that $\Omega_n$ depends only on the system $\{\omega^{(g)}_n\}$ and not on the choice
of~$x$ and~$y$.

\begin{definition}[{\cite[Def.~3.2]{ABDKS05}}]\label{def:bW}
The \emph{extended $n$-point differential} is obtained from $\Omega_n$ by
moving the two legs of each variable onto a curve through the diagonal,
\begin{align}\label{eq:bW-graph}
  \bW_n(p_{\set n},u_{\set n})
  =\prod_{i=1}^n\Bigl(e^{u_i(\cS(u_i\hbar\partial_{x_i})-1)y_i}
   \Bigr)\,
   \Omega_n(p^+_{\set n},p^-_{\set n}) \Big|_{\forall i\, p^\pm_i= e^{\pm\frac{u_i\hbar}2\partial_{x_i}}p_i},
  \qquad
  \cS(u)=\frac{e^{u/2}-e^{-u/2}}{u}.
\end{align}
It is a symmetric $n$-differential in the $n$ pairs $(p_i,u_i)$. By $e^{u\partial_x}p$ we mean here the phase trajectory of the vector field $\partial_x$ issued from $p$. Equivalently, this specialization corresponds to the substitution $x(p^\pm_i)=x_i\pm\tfrac{u_i\hbar}2$.
\end{definition}

Setting all the $u_i$ to zero brings the two legs of every variable
together and recovers the correlators,
\begin{align}\label{eq:bW-alpha-zero}
  \coeff{u_{\set n}}{0}\,\bW_n=\omega_n .
\end{align}
Here $u$, and later $v$, are formal parameters.

The following property of these differentials is important: each coefficient of a power of
$\hbar$ in $\bW_n-\delta_{n,1}dx_1/(u_1\hbar)$ is a polynomial in the
$u_i$, so that $\bW^{(g)}_n$ is a differential polynomial in the
$\omega^{(g')}_{n'}$; no integration is actually performed. This fact can also be expressed by the following explicit formula:
\begin{align}\label{eq:bW-explicit}
  \bW_n & =\prod_{i=1}^n\Bigl(\frac{dx_i}{u_i\hbar}\,
    e^{u_i(\cS(u_i\hbar\partial_{x_i})-1)y_i}\Bigr)
  \sum_{\Gamma\ \mathrm{connected}}\frac{1}{|\Aut\Gamma|}
 \\ \notag & \quad \times
  \prod_{e\in E(\Gamma)} \left(\prod_{j=1}^{|e|}
   \Bigl(
    u_{e(j)}\hbar\,\cS\bigl(u_{e(j)}\hbar\,
      \partial_{\widetilde x_j}\bigr)\Bigr)
   \frac{\widetilde\omega_{|e|}(\widetilde p_{\set{|e|}})}
        {\prod_{j}d\widetilde x_j}\right)\Big|_{\forall j\,\widetilde p_j=p_{e(j)}}.
\end{align}
Here the normalization of $\widetilde\omega$ is fixed with $\frac{dx_1 dx_2}{(x_1-x_2)^2}$ (that is, in Equation~\eqref{eq:regularised-om02} the function $z$, which is arbitrary there, is fixed to be $x$ here).

\begin{remark}\label{rem:x-usage}
The above formulas involve only $dx$ and $dy$, not the functions $x$ and $y$ themselves. Therefore, for meromorphy of $\bW_n$ it is sufficient to assume that $dx$ and $dy$ are meromorphic.
\end{remark}

\subsection{Definition of \texorpdfstring{$\xy$}{x-y} swap and symplectic duality transformations}

\begin{definition}\label{def:xy-dual}
The $\xy$ dual system of differentials is defined by
\begin{equation}\label{eq:xy-dual-def}
\omega_n^\vee=\prod_{i=1}^n\left(\sum_{r\ge0}
\bigl(-d_i\tfrac{1}{dy_i}\bigr)^r\coeff{u_i}{r}\right)\bW_n.
\end{equation}
\end{definition}

\begin{definition}\label{def:sympl-dual}
For given symplectic duality data $(x,y,\widetilde\psi)$, we set
\begin{equation}\label{eq:x-dagger-def}
x^\dagger(p)=x(p)+\psi(y(p)).
\end{equation}
The symplectic dual system of differentials is defined by
\begin{align}\label{eq:sympl-dual-def}
\omega_n^\dagger & =\prod_{i=1}^n\left(\sum_{r,j\ge0}
\bigl(-d_i\tfrac{1}{dx^\dagger_i}\bigr)^j\coeff{v}{j}
\biggl(\frac{\partial_{y}^re^{v\cS(v\hbar\partial_y)\widetilde\psi(y)}}
{e^{v\psi(y)}}\bigg|_{y=y_i}\biggr)
\coeff{u_i}{r}\right)\bW_n
\\ \notag & \quad -\delta_{n,1}\hbar^{-1}\sum_{j\ge1}
\bigl(-d_1\tfrac{1}{dx^\dagger_1}\bigr)^{j-1}\coeff{v}{j}
\biggl(\frac{e^{v\cS(v\hbar\partial_y)\widetilde\psi(y)}}
{e^{v\psi(y)}}\bigg|_{y=y_1}\biggr)dy_1.
\end{align}
\end{definition}

\begin{theorem}\label{thm:xy-sympl-properties}
1. The differentials of both systems $\omega^\vee_n$ and $\omega^\dagger_n$ are symmetric, possess the topological expansion property, and are regular on the diagonals except for the standard diagonal singularity of the $(0,2)$ differential
\begin{equation}\label{eq:dual-02-differentials}
\omega^{\vee,(0)}_2=\omega^{\dagger,(0)}_2=\omega^{(0)}_2.
\end{equation}

2. The inverse transformations of the $\xy$ swap and symplectic duality are also $\xy$ swap and symplectic duality transformations for the data $(x^\vee,y^\vee)=(y,-x)$ (or $(-y,x)$, since changing the signs of both $x$ and $y$ defines the same transformation) and $(x^\dagger,y^\dagger,\widetilde\psi^\dagger)=(x+\psi(y),y,-\widetilde\psi)$, respectively.

3. The $\xy$ dual differentials $\omega^{\vee,(g)}_n$ of the original system with respect to the given functions $(x,y)$ and the $\xy$ dual differentials $\omega^{\vee,\dagger,(g)}_n$ of the symplectic dual system with respect to the functions $(x^\dagger,y)$ are related by
\begin{equation}\label{eq:xy-vs-sympl-composition}
\omega^{\vee,\dagger}_n=\omega^{\vee}_n-\delta_{n,1}\hbar^{-1}(\widetilde\psi(y_1)-\psi(y_1))\,dy_1.
\end{equation}
In other words, symplectic duality can be realized as a composition of three transformations: 1) $\xy$ swap for the original functions $(x,y)$; 2) modification of the $1$-point differential by subtracting $\hbar^{-1}(\widetilde\psi(y_1)-\psi(y_1))\,dy_1$; 3) one more $\xy$ swap, this time with $(y,-x^\dagger)$ as the $(x,y)$ functions.
\end{theorem}

Above, symplectic duality is used in the formulation of topological recursion. Let us give an explicit formula for $\cW^{(g)}_{n}(v,p)$ used in Equations~\eqref{eq:mixed-as-differential-polynomial} and~\eqref{eq:tildeomega-def}:
\begin{align} \label{eq:curly-W-definition}
	\cW^{(g)}_{n}(v,p) & \coloneqq [\hbar^{2g-2+n}]\left( \sum_{r\ge 0}
	\biggl(\frac{\partial_{y}^re^{v\cS(v\hbar\partial_y)\widetilde\psi(y)}}
	{e^{v\psi(y)}}\biggr)
	\coeff{u}{r}\bW_n\Big|_{\substack{u_1=\cdots=u_{n-1}=0 \\ u_n=u, p_n=p}}
-\delta_{n,1}
	\biggl(\frac{e^{v\cS(v\hbar\partial_y)\widetilde\psi(y)}}
	{e^{v\psi(y)}}\biggr)\frac{dy}{v\hbar}\right),
\end{align}
where $y=y(p)$.

\subsection{Transformations for extended differentials}

There are obvious combinatorial difficulties in handling formulas for the $\xy$ swap and symplectic duality transformations. However, the situation simplifies dramatically when these transformations are applied to the \emph{extended} $\Omega$-half-differentials: both are represented as products of $2n$ integral transformations acting independently on the $2n$ arguments.

For given two functions  $x$, $y$ on the spectral curve, denote by $G_{x,y}$ the transformation taking a function $f$ on the spectral curve to the function
\begin{equation}\label{eq:Gxy-appendix}
(G_{x,y}f)(q)=\frac{1}{\sqrt{2\pi\hbar}}\int f(p)\;
 e^{\frac{1}{\hbar}\bigl((x(q)-x(p))\,y(q)+\int\limits_q^py\,d x\bigr)}\dd x(p).
\end{equation}
This transformation acts by a differential operator expanded in $\hbar$ and with coefficients depending on the derivatives of $x$ and $y$:
\begin{equation}\label{eq:Gxy-operator-expansion}
G_{x,y}f=\frac{1}{\sqrt{y'}}\biggl(
f+\Bigl(\bigl(\tfrac{y^{(3)}}{8\,y'^2}-\tfrac{5\,y''^2}{24\,y'^3}\bigr)\,f+
\tfrac{y''}{2\,y'^2}\,f'-\tfrac{1}{2\,y'}\,f''\Bigr)\,\hbar+O(\hbar^2)\biggr),
\end{equation}
where the prime denotes $\frac{d}{dx}$. The presence of the radical factor in the last equality suggests that it is more natural to regard this operator acting in the space of half-differentials rather than functions. Namely, set
\begin{equation}\label{eq:bGxy-half-differential}
\bG_{x,y}(f\sqrt{dx})=(G_{x,y}f)\sqrt{dy}.
\end{equation}
Then, for any half-differential $\phi$, we have
\begin{equation}\label{eq:bG-leading-term}
\bG_{x,y}\phi=\phi+O(\hbar),
\end{equation}
and this equality fixes the sign ambiguity in the Gaussian integral involved in the definition of this transformation. The transformation $\bG_{x,y}$ is invertible and the inverse one is given by
\begin{equation}\label{eq:bG-inverse}
\bG_{x,y}^{-1}=\bG_{y,-x}=\bG_{-y,x}.
\end{equation}
The two forms of the inverse transform are equivalent since the change of the signs of both $x$ and $y$ preserves $\bG_{x,y}$.

With this notation, the $\xy$ swap transformation acts on $\bK=\Omega_1$ by
\begin{equation}\label{eq:bK-xy-dual}
\bK^\vee(p_1,p_2)= \bG^{(p_1)}_{x,y}\;\bG^{(p_2)}_{x,-y} \bK(p_1,p_2).
\end{equation}
In a similar way, for any $n\ge1$ we have
\begin{equation}\label{eq:Omega-xy-dual}
\Omega^\vee_n(p^+_{\set n},p^-_{\set n})
=\left(\prod_{i=1}^n \bG^{(p_i^+)}_{x,y}\;\bG^{(p_i^-)}_{x,-y}\right)~
\Omega_n(p^+_{\set n},p^-_{\set n}).
\end{equation}

As an immediate corollary, we conclude that the $\xy$ swap transformation preserves KP integrability (just apply
$\prod_{i=1}^n \bG^{(p_i^+)}_{x,y}\;\bG^{(p_i^-)}_{x,-y}$ to both sides of the determinantal identity). Similarly, it preserves BKP integrability as well (under extra assumptions --- for the BKP integrability we assume that the curve is equipped with an involution $\iota$ involved in the definition of BKP integrability and that $x$ and $y$ have opposite parities with respect to~$\iota$), which we discuss in the next Section.
\begin{remark}\label{rem:bG-as-Sato-group-element}
Moreover, the operator $\bG_{x,y}$ can often be identified with Sato’s group element, acting on the corresponding version of the Sato Grassmannian~\cite[Section 2]{A_H32}.
\end{remark}

Since symplectic duality decomposes as a composition of two $\xy$ swap transformations and a $1$-point correction, it has a similar integral representation:
\begin{equation}\label{eq:sympldual-Omega}
\Omega^\dagger_n(p^+_{\set n},p^-_{\set n})
=\left(\prod_{i=1}^n \bG^{(p_i^+)}_{x,y,\widetilde\psi}\;\bG^{(p_i^-)}_{-x,y,-\widetilde\psi}\right)~
\Omega_n(p^+_{\set n},p^-_{\set n}),
\end{equation}
where $\bG_{x,y,\widetilde\psi}:f\sqrt{dx}\mapsto f^\dagger\sqrt{dx^\dagger}$, $x^\dagger=x+\psi(y)$, is given by the double Gaussian integral
\begin{equation}\label{eq:bG-sympl-integral}
\begin{aligned}
f^\dagger(q)&=\frac{i}{2\pi\hbar}\iint f(p_1)\;
 e^{\frac{1}{\hbar}\bigl((x(p_2)-x(p_1))\,y(p_2)+\int\limits_{p_2}^{p_1}y\,d x\bigr)}\dd x(p_1)\;
\\&\hskip3cm \times e^{-\frac{1}{\hbar}\int_{p_1}^{p_2}(\widetilde\psi(y)-\psi(y))\,dy}\;
\\&\hskip3cm \times e^{-\frac{1}{\hbar}\bigl((y(q)-y(p_2))\,x^\dagger(q)+\int\limits_{q}^{p_2}x^\dagger \,d y\bigr)}\dd y(p_2).
\end{aligned}
\end{equation}

Symplectic duality also preserves KP and BKP integrabilities (the KP case is discussed in~\cite{ABDKS05}, and the BKP case we discuss in the next Section), and the corresponding transformation of the $K$-kernel is given by~\eqref{eq:sympldual-Omega} for~$n=1$.



\subsection{Global BKP integrability}\label{sec:global-BKP}
We describe the BKP analogue of global KP integrability, with the normalization of~\eqref{eq:omega}--\eqref{eq:detid-rescaled}.
Let $\Sigma$ be a connected Riemann surface with a holomorphic involution $\iota\not=\mathrm{Id}_\Sigma$, $\iota^2=\mathrm{id}$, and let $\{\omega_n\}_{n\geq1}$ be symmetric meromorphic $n$-differentials on $\Sigma^n$, odd in each argument:
\begin{equation}\label{eq:global-BKP-odd}
	\iota_i^*\omega_n=-\omega_n,\qquad i=1,\dots,n,
\end{equation}
where by  $\iota_i$ we denote the action of $\iota$ on the $i$-th copy of $\Sigma$ in $\Sigma^n$.

We assume that $\omega_2$ has a double pole with biresidue~$1$ along the diagonal in $\Sigma^2$, and that the other differentials are regular on all partial diagonals.

Consider a simply-connected domain $\tilde{\Sigma}\subseteq\Sigma$ 
such that $\iota$ can be restricted to it. Define on $\tilde{\Sigma}^2$
the bi-half-differential
\begin{equation}\label{eq:Omega1}
	\bK(p,q)=\frac{\sqrt{d\chi(p)}\sqrt{d\chi(q)}}{\chi(p)-\chi(q)}
	\exp\left(\sum_{m\geq1}\frac1{m!}
	\Bigl(\int_q^p\Bigr)^m\left(\omega_m-\delta_{m,2}\frac{d\chi(p)\,d\chi(q)}{(\chi(p)-\chi(q))^2}\right)\right),
\end{equation}
where $\chi$ is an arbitrary local coordinate, the definition does not depend on this choice. Here the initial definition is near a generic point of the diagonal, with short integration paths. At a fixed point of the involution this means a punctured neighborhood: in the formula we removed the diagonal pole but not the antidiagonal. Extending further to $\tilde{\Sigma}$ might require a choice of cuts to fix the branches of logarithm at the poles of $\omega_m$.

\begin{definition}\label{def:global-BKP}
	The system of differentials $\{\omega_n\}_{n\geq 1}$ is \emph{globally BKP integrable} if
	\begin{equation}\label{eq:global-BKP-determinant}
		\omega_n(p_{\set n})=
		{\rm Alt}_1\cdots{\rm Alt}_{n-1}(-1)^{n-1}
		\sum_{\sigma\in\operatorname{cycl}(n)}
		\prod_{i=1}^n\bK(p_i,p_{\sigma(i)}),\qquad n\geq2,
	\end{equation}
	where ${\rm Alt}_i=\mathrm{id}-\iota_i^*$, acting on $i$-th arguments of the differentials.
\end{definition}

The one-point identity
\begin{equation}\label{eq:global-BKP-one-point}
	\omega_1(p)=\left.\Bigl(\bK(p,q)-\frac{\sqrt{d\chi(p)}\sqrt{d\chi(q)}}{\chi(p)-\chi(q)}
	\Bigr)\right|_{q=p}
\end{equation}
holds by construction (for an arbitrary choice of $\chi$), without an integrability assumption. Also, oddness gives
\begin{equation}\label{eq:global-BKP-kernel-symmetry}
	(\iota\times\iota)^*\bK(p,q)=-\bK(q,p).
\end{equation}
Consequently, the right-hand side of~\eqref{eq:global-BKP-determinant} is already odd in the last variable. As we mentioned above, equivalently, one may apply all $n$ alternations and insert an overall factor~$1/2$.

\subsubsection{Regular fixed point} In order to pass from a system of differentials to a partition function, we need to choose a point of expansion on $\Sigma$ and a local coordinate at this point.

\begin{definition}\label{def:BKP-regular-point}
	A fixed point $o$ of $\iota$ is called \emph{regular} for the system if, in an odd local coordinate $z$, $z(o)=0$, $z\circ\iota=-z$, the differentials
	\begin{equation}\label{eq:BKP-regular-remainder}
		\omega_n(p_{\set{n}})-\delta_{n,2}\left( \frac{dz(p_1) dz(p_2)}{(z(p_1)-z(p_2))^2}
		+\frac{dz(p_1)dz(p_2)}{(z(p_1)+z(p_2))^2}\right),
	\end{equation}
	are holomorphic at $o^n$, for every $n\geq1$.
\end{definition}

\begin{remark}\label{rem:odd-coordinate-exists}
	An odd coordinate exists at every fixed point of $\iota$ and this definition is independent of this coordinate. We assume throughout this subsection that at least one regular fixed point exists, and fix such a point $o$. Its existence is an additional hypothesis on $(\Sigma,\iota,\{\omega_n\})$. Part of the theory can be built up without this assumption, but first, some proofs do require it, and second, it is needed in any case if we want to extract some enumerative data from $\{\omega_n\}_{n\geq 1}$.
\end{remark}

At a regular fixed point, introduce the coefficients $C_{k_1,\dots,k_n}$ and the associated normalized series $\tau_{o,z}$ by
\begin{align}
	& \omega_n=\delta_{n,2}\left( \frac{dz_1 dz_2}{(z_1-z_2)^2}
	+\frac{dz_1dz_2}{(z_1+z_2)^2}\right)+
	\sum_{\substack{k_1,\dots,k_n>0\\k_i\ \odd}}
	C_{k_1,\dots,k_n}\prod_{i=1}^n d(z_i^{k_i}),
	\label{eq:BKP-local-expansion}\\
	& \log\tau_{o,z}(p_1,p_3,\dots)
	=\sum_{n\geq1}\frac1{n!}
	\sum_{\substack{k_1,\dots,k_n>0\\k_i\ \odd}}
	C_{k_1,\dots,k_n}\prod_{i=1}^n p_{k_i}.
	\label{eq: tau-BKP}
\end{align}
Then we obtain directly from~\eqref{eq:Omega1} the following expansion for $\bK$ (assuming that $o\in \tilde\Sigma$, or, otherwise, we would have to discuss the global extension of $\bK$ and related subtleties):
\begin{equation}\label{eq:BKP-local-kernel}
	\bK(p,q)=\frac12\frac{z(p)+z(q)}{z(p)-z(q)}
	\sqrt{\frac{dz(p)}{z(p)}}\sqrt{\frac{dz(q)}{z(q)}}
	\tau_{o,z}\big|_{p_k=z(p)^k-z(q)^k}.
\end{equation}

\begin{proposition}\label{prop:global-BKP-local}
	The system of differentials $\{\omega_n\}_{n\geq 1}$ is globally BKP integrable if and only if $\tau_{o,z}$ is a BKP tau function for one, hence every, odd local coordinate $z$ at any regular fixed point $o$.
\end{proposition}

\begin{proof}
	At the point $o$,~\eqref{eq:BKP-local-kernel} identifies $\bK$ with the neutral-fermion two-point function in the normalization of~\eqref{eq:BA}. The connected neutral Wick formula is exactly~\eqref{eq:global-BKP-determinant}; its one-point version is~\eqref{eq:global-BKP-one-point}, see~\cite{WY}. Conversely, the skew affine coordinates read from this two-point function define a normalized BKP tau function by~\eqref{eq:BKP-Pfaffian-expansion}. Wick's formula and the assumed identities give it the same connected correlators as $\tau_{o,z}$. Hence their logarithms have identical Taylor expansions, so the two series coincide. Coordinate independence and analytic continuation give the asserted global identities. Expanding them in another odd coordinate, or at another fixed point, gives the same local criterion.
\end{proof}

\begin{remark}\label{rem:several-regular-fixed-points}
In the case when there are several regular fixed points one can construct the multi-component BKP tau functions in a way, similar to~\cite[Section 2.8]{ABDKS10}.
Moreover, we expect that expansions at the $\iota$ invariant pairs of non-fixed regular points also can be described by an integrable hierarchy in a way, similar to one considered in \cite{KrZ}.
\end{remark}

\subsubsection{Extended differentials}

We extend the concept of global BKP integrability defining the extended $n|n$ half-dif\-fe\-ren\-tials. Let $B$ be an auxiliary symmetric bidifferential with the usual diagonal double pole of biresidue~$1$; it need not be alternated. Denote by $E_B$ its prime form defined as
\begin{equation}\label{eq:BKP-prime-factor}
	\frac1{E_B(p,q)}=\frac{\sqrt{d\chi(p)}\sqrt{d\chi(q)}}{\chi(p)-\chi(q)}
	\exp\left(\frac12\int_q^p\int_q^p\left(B-\frac{d\chi(p)\,d\chi(q)}{(\chi(p)-\chi(q))^2}\right)\right).
\end{equation}
As above, the right hand side of this formula does not depend on the choice of $\chi$. The prime form satisfies $d_p d_q\log E_B(p,q)=B(p,q)$.

The disconnected extended $n|n$ half-differentials are defined as
\begin{align}\label{eq:Omega-bullet-expint}
	\Omega_n^\bullet(p^+_{\set n},p^-_{\set n})
	& \coloneqq \prod_{i=1}^n\frac1{E_B(p_i^+,p_i^-)}
	\prod_{i<j}
	\frac{E_B(p_i^+,p_j^+)E_B(p_i^-,p_j^-)}
	{E_B(p_i^+,p_j^-)E_B(p_i^-,p_j^+)}
	\nonumber\\
	&\quad\times\exp\left(
	\sum_{m\geq1}\frac1{m!}
	\left(\sum_{i=1}^n \int_{p_i^-}^{p_i^+}\right)^m(\omega_m-\delta_{m,2}B)
	\right).
\end{align}
The power of the sum of integration operators means integration in $m$ arguments. The connected extended $n|n$ half-differentials are defined by
\begin{equation}\label{eq:Omega-expint}
	\Omega_n(p^+_{\set n},p^-_{\set n})=
	\sum_{\pi\in\mathcal P(\set n)}
	(-1)^{|\pi|-1}(|\pi|-1)!
	\prod_{J\in\pi}\Omega_{|J|}^\bullet(p_J^+,p_J^-),
\end{equation}
where $\mathcal P(\set n)$ is the set of unordered set partitions. For the proof of the following Lemma see e.~g.,~\cite{ABDKS03} or~\cite[Section 1.2]{ABDKS08}:

\begin{lemma}\label{lem:BKP-extended-definition}
	These definitions are independent of $B$, and $\Omega_1=\Omega_1^\bullet=\bK$. They agree with the graph definition~\eqref{eq:Omega-graph}. For $n\geq2$ they have no pole on a generic divisor $p_i^+=p_i^-$, and
	\begin{equation}\label{eq:BKP-extended-diagonal}
		\Omega_n(p^+_{\set n},p^-_{\set n})\big|_{p_i^+=p_i^-=p_i}
		=\omega_n(p_{\set n}).
	\end{equation}
\end{lemma}

Define the extension of ${\rm Alt}_i$ to pairs of arguments by applying both the involution and the exchange of the points:
\begin{equation}\label{eq:BKP-paired-alternation}
	J_i\colon(p_i^+,p_i^-)\mapsto(\iota(p_i^-),\iota(p_i^+)),
	\qquad {\rm Alt}_i^\pm=\mathrm{id}-J_i^*.
\end{equation}
Pullback includes the half-differentials, with the choices of branches being fixed in advance. On the paired diagonal this reduces to ${\rm Alt}_i$ acting on differentials. Near a regular fixed point, in the invariant frames $\sqrt{dz/z}$, it reads
\begin{equation}\label{eq:BKP-paired-alternation-local}
	({\rm Alt}_i^\pm f)(z_i^+,z_i^-)
	=f(z_i^+,z_i^-)-f(-z_i^-,-z_i^+).
\end{equation}

\begin{proposition}\label{prop:equiv-KP}
	Under the assumption of existence of a regular fixed point, the global BKP integrability is equivalent to the extended identities
	\begin{equation}\label{eq:global-BKP-extended-determinant}
		\Omega_n(p^+_{\set n},p^-_{\set n})=
		{\rm Alt}_1^\pm\cdots{\rm Alt}_{n-1}^\pm(-1)^{n-1}
		\sum_{\sigma\in\operatorname{cycl}(n)}
		\prod_{i=1}^n\bK(p_i^+,p_{\sigma(i)}^-),\qquad n\geq1.
	\end{equation}
	For $n=1$ this means $\Omega_1=\bK$. For $n\geq2$ the right-hand side is already $J_n$-odd; applying the last alternation would require a factor~$1/2$.
\end{proposition}

\begin{proof} It is an adaptation of the proof of~\cite[Theorem 4.4]{ABDKS03} to the context of BKP integrability. For $n=1$, the identity is tautological. For $n\geq 2$, we proceed with a formal expansion and use tools from BKP integrability.
	
	Use the chosen regular fixed point $o$ and an odd coordinate there. By Proposition~\ref{prop:global-BKP-local}, the local series is a BKP tau function. In the neutral-fermion setup~\cite{DJKM}, see also~\cite{WY}, let $|V\rangle$ be its normalized state, so that
	\begin{align}
	\tau_{o,z}(p)=\langle0|e^{\sum_{k>0,\,k\,\odd}(2p_k/k)H_k^B}|V\rangle,
	\qquad \langle0|V\rangle=1.\label{eq:BKP-tau-from-neutral-state}
	\end{align}
	The bosonization identifies~\eqref{eq:Omega-bullet-expint} with
	\begin{equation}\label{eq:BKP-neutral-expectation}
		\Omega_n^\bullet=
		\prod_i\sqrt{\frac{dz_i^+}{z_i^+}}\sqrt{\frac{dz_i^-}{z_i^-}}\,
		\langle0|\prod_{i=1}^n
		\bigl(\phi(1/z_i^-)\phi(-1/z_i^+)\bigr)|V\rangle.
	\end{equation}
	Indeed, its bosonic expression is
	\begin{align}\label{eq:BKP-local-disconnected}
		\Omega_n^\bullet & =
		\prod_i\left(\frac{z_i^++z_i^-}{2(z_i^+-z_i^-)}
		\sqrt{\frac{dz_i^+}{z_i^+}}\sqrt{\frac{dz_i^-}{z_i^-}}\right)
		\\ \notag & \quad
		\times\prod_{i<j}
		\frac{(z_i^+-z_j^+)(z_i^--z_j^-)(z_i^++z_j^-)(z_i^-+z_j^+)}
		{(z_i^+-z^-_j)(z^-_i-z_j^+)(z^+_i+z_j^+)(z^-_i+z^-_j)}
		\left.\tau_{o,z}\right|_{p_k=\sum_i((z_i^+)^k-(z_i^-)^k)}.
	\end{align}
	Neutral Wick's formula~\cite{WY} computes~\eqref{eq:BKP-neutral-expectation} as a Pfaffian of the $2n\times2n$ skew-symmetric matrix with entries $\bK(q_\alpha,-q_\beta)$, where $q_{2i-1}=z_i^+$ and $q_{2i}=- z_i^-$. Taking its connected part retains the pairings forming a single cycle through the $n$ pairs of variables $(z_i^+,z_i^-)$.
	For each $\sigma\in\operatorname{cycl}(n)$ we have the term $(-1)^{n-1}
	\prod_{i=1}^n\bK(z_i^+,z_{\sigma(i)}^-)$ and all further terms that can be obtained from this one by transformations $(z_i^+,z_i^-)\to(-z_i^-,-z_i^+)$, $i=1,\dots,n-1$, where each such transformation also adds a factor of $(-1)$. This gives precisely~\eqref{eq:global-BKP-extended-determinant}, expanded locally near $o$. The identity then analytically continues globally.
	
	Conversely, taking the paired diagonal in~\eqref{eq:global-BKP-extended-determinant} and using~\eqref{eq:BKP-extended-diagonal} gives~\eqref{eq:global-BKP-determinant}.
\end{proof}

\subsubsection{BKP integrability and duality transformations}\label{sec:BKP-preservation}

In this section, we discuss how the $\xy$ swap and symplectic duality preserve global BKP integrability. To this end, we adapt the arguments of~\cite[Theorems 4.3--4.4]{ABDKS03} and~\cite[Theorems 5.1--5.3]{ABDKS05} to the determinantal identities~\eqref{eq:global-BKP-extended-determinant}.

\begin{theorem}\label{thm:BKP-preservation}
	Let $\{\omega_n\}$ be globally BKP integrable on $(\Sigma,\iota)$. Consider $x$ and $y$, or rather $dx$ and $dy$, such that
	\begin{equation}\label{eq:BKP-swap-parities}
		\iota^*dx=dx,\
		\iota^*dy=-dy,\qquad  \text{or} \qquad
		\iota^*dx=-dx,\
		\iota^*dy=dy.
	\end{equation}
	Then the $\xy$ swap preserves the extended BKP determinantal identities. The same holds for symplectic duality if
	\begin{equation}\label{eq:BKP-sympl-parities}
		\iota^*dx=dx, \ \iota^*y=-y,\ \text{and }\
		\widetilde\psi(-\theta) = \widetilde\psi(\theta),
	\end{equation}
	or, more generally, $\psi'(-\theta) = -\psi'(\theta)$ and $\psi_k(-\theta) = \psi_k(\theta)$, $k\geq 1$.
	
	More explicitly, we have, respectively,
	\begin{align}\label{eq:BKP-dual-extended-determinant}
		\Omega_n^\vee(p^+_{\set n},p^-_{\set n})& =
		{\rm Alt}_1^\pm\cdots{\rm Alt}_{n-1}^\pm(-1)^{n-1}
		\sum_{\sigma\in\operatorname{cycl}(n)}
		\prod_{i=1}^n\bK^\vee(p_i^+,p_{\sigma(i)}^-),\qquad n\geq1.
		\\
		\Omega_n^\dagger(p^+_{\set n},p^-_{\set n}) & =
		{\rm Alt}_1^\pm\cdots{\rm Alt}_{n-1}^\pm(-1)^{n-1}
		\sum_{\sigma\in\operatorname{cycl}(n)}
		\prod_{i=1}^n\bK^\dagger(p_i^+,p_{\sigma(i)}^-),\qquad n\geq1.
	\end{align}
	
	Consequently, whenever the transformed system of differentials has a regular fixed point, it is globally BKP integrable. If both systems admit such points, then the BKP integrability holds on one side of either duality if and only if it holds on the other side.
\end{theorem}

\begin{proof} In the case of $\xy$ swap,
	the integral transformations depend only on $dx$ and $dy$.
	In both cases in~\eqref{eq:BKP-swap-parities}, changing the integration variable by $\iota$ in~\eqref{eq:Gxy-appendix} gives
	\begin{equation}\label{eq:BKP-G-conjugacy}
		\iota^*\bG_{x,y}=\bG_{x,-y}\iota^*,\qquad
		\iota^*\bG_{x,-y}=\bG_{x,y}\iota^*.
	\end{equation}
	Here the signs of the half-differentials are fixed by~\eqref{eq:bG-leading-term} and we also use $\bG_{-x,-y}=\bG_{x,y}$. Hence for  $J_i\colon (p_i^+,p_i^-)\mapsto(\iota(p_i^-),\iota(p_i^+))$ (defined above in~\eqref{eq:BKP-paired-alternation})
	we have
	\begin{equation}\label{eq:BKP-paired-transform-commutation}
		\Big(\prod_{k=1}^n\bG_{x,y}^{(p_k^+)}\bG_{x,-y}^{(p_k^-)}\Big)\, J_i^*
		=J_i^*	\prod_{k=1}^n\bG_{x,y}^{(p_k^+)}\bG_{x,-y}^{(p_k^-)}, \qquad i=1,\dots,n.
	\end{equation}
	Apply $\prod_{i=1}^n\bG_{x,y}^{(p_i^+)}\bG_{x,-y}^{(p_i^-)}$ to~\eqref{eq:global-BKP-extended-determinant}. By~\eqref{eq:Omega-xy-dual}, the left-hand side becomes $\Omega_n^\vee$. The operator commutes with every paired alternation, and each endpoint occurs in exactly one kernel of each cyclic product. Thus~\eqref{eq:bK-xy-dual} transforms that product into the corresponding product of $\bK^\vee$, proving~\eqref{eq:BKP-dual-extended-determinant} for the $\xy$ swap. Restriction to the paired diagonal gives~\eqref{eq:global-BKP-determinant}. The $J_i$-oddness, including the symmetry of $\bK$ given in~\eqref{eq:global-BKP-kernel-symmetry}, is preserved as well, so the resulting differentials are odd in each argument.
	
	For symplectic duality, we use~\eqref{eq:xy-vs-sympl-composition}. The intermediate modification on the dual side is $\omega_1^\vee\mapsto\omega_1^\vee+\eta$, where
	\begin{equation}\label{eq:BKP-sympl-one-point-shift}
		\eta=-\frac 1\hbar (\widetilde\psi(y)-\psi(y))\,dy
	\end{equation}
	For any system, adding $\eta$ to its one-point differential changes the kernel and extended differentials by
	\begin{equation}\label{eq:BKP-odd-gauge}
		{\bK}(p,q) \to e^{\int_q^p\eta}\bK(p,q),
		\qquad
		\Omega_n(p^+_{\set n},p^-_{\set n}) \to \left(\prod_{i=1}^n
		e^{\int_{p_i^-}^{p_i^+}\eta}\right)\Omega_n(p^+_{\set n},p^-_{\set n}).
	\end{equation}
	Since $\eta$ is odd, each factor in this product is invariant under $J_i$. A cyclic product of the kernels in \eqref{eq:global-BKP-extended-determinant} acquires precisely the same factor. Thus the intermediate modification on the dual side preserves the extended determinantal identities. Combining this with the statement on $\xy$ swap completes the proof.
\end{proof}

\begin{remark}\label{rem:regular-fixed-point-needed}
	The proof preserves the extended determinantal identities without requiring a regular fixed point for the intermediate or final systems. Such a point is needed to pass from the usual determinantal identities to extended determinantal identities (at least, we do not have a proof without this assumption). Hence, for either transformation, the BKP integrability of $\{\omega_n\}_{n\geq 1}$ implies the BKP integrability of the transformed differentials only under this extra assumption.
\end{remark}

\begin{remark}\label{rem:no-topological-expansion-needed} Note that we have not stated explicitly any assumption on topological expansion of $\{\omega_n\}_{n\geq 1}$ in $\hbar$ in the statement of Theorem~\ref{thm:BKP-preservation}. It is not used in the proof, and in fact this theorem can be applied in a more general situation, cf. e.g.,~\cite{ABDKS08} in the KP context.
\end{remark}

\begin{remark}\label{rem:parities-in-this-paper}
	In the situation used in this paper, $\iota^*dx=dx$ and $y\circ\iota=-y$. The symplectic assumptions are then satisfied if
	\[
	\psi'(-\theta)=-\psi'(\theta),\qquad
	\psi_k(-\theta)=\psi_k(\theta),\quad k\geq1.
	\]
	The condition on $\psi'$ allows logarithmic functions, in particular $\psi(\theta)=\log\theta^2$, as above.
\end{remark}



\subsection{Mixed \texorpdfstring{$(k,n-k)$}{(k,n-k)} differentials} \label{sec:mixed-differentials}
The study of the $\xy$ swap and symplectic duality in~\cite{ABDKS01,BDKS04,ABDKS05} prompts a definition of the mixed $(k,n-k)$ differentials, especially natural in the context of the two-matrix model. Let us record some potentially useful formulas for the mixed differentials in the BKP context. We do this for the $\xy$ swap; for the symplectic duality the description is completely parallel.

The extended $(k,n-k)$ (half-)differentials are defined as
	\begin{align} \label{eq: mixed Omega}
	\Omega_{k,n-k}(p^+_{\set n},p^-_{\set n})& = \Big(\prod_{i=k+1}^n\bG_{x,y}^{(p_i^+)}\bG_{x,-y}^{(p_i^-)}\Big)\Omega_{n}(p^+_{\set n},p^-_{\set n}).
\end{align}
Thus, $\Omega_{n}=\Omega_{n,0}$ and $\Omega_n^\vee = \Omega_{0,n}$. The mixed $(k,n-k)$ differentials are obtained by restriction of~\eqref{eq: mixed Omega} to the paired diagonals:
\begin{align}
		\omega_{k,n-k}(p_{\set n})& =
		 \left.\Big(\prod_{i=k+1}^n\bG_{x,y}^{(p_i^+)}\bG_{x,-y}^{(p_i^-)}\Big)\Omega_{n}(p^+_{\set n},p^-_{\set n})\right|_{p_i^+=p_i^-=p_i}, \qquad n\geq 2;\label{eq:mixed-omega-def}
\end{align}
for $n=1$ one has to apply a regularization as in~\eqref{eq:global-BKP-one-point}.

By virtue of the determinantal formulas~\eqref{eq:global-BKP-determinant},~\eqref{eq:global-BKP-extended-determinant}, we obtain the following determinantal formulas for the mixed differentials:
	\begin{equation}
	\Omega_{k,n-k}(p^+_{\set n},p^-_{\set n})=
	{\rm Alt}_1^\pm\cdots{\rm Alt}_{n-1}^\pm(-1)^{n-1}
	\sum_{\sigma\in\operatorname{cycl}(n)}
	\prod_{i=1}^n\widehat{\bK}(p_i^+,p_{\sigma(i)}^-),\qquad n\geq1,\label{eq:mixed-extended-determinant}
\end{equation}
where
\begin{align}\label{eq:mixed-kernel-cases-extended}
	\widehat{\bK}(p^+_i,p^-_{\sigma(i)}) = \begin{cases}
		{\bK}(p^+_i,p^-_{\sigma(i)}) & i, \sigma(i) =1,\dots,k; \\
		\bG_{x,y}^{(p^+_i)} {\bK}(p^+_i,p^-_{\sigma(i)}) & i=k+1,\dots,n,\ \sigma(i) =1,\dots,k; \\
		\bG_{x,-y}^{(p^-_{\sigma(i)})} {\bK}(p^+_i,p^-_{\sigma(i)}) & i=1,\dots,k,\ \sigma(i) =k+1,\dots,n; \\
		\bG_{x,y}^{(p^+_i)}\bG_{x,-y}^{(p^-_{\sigma(i)})} {\bK}(p^+_i,p^-_{\sigma(i)}) & i,\sigma(i) =k+1,\dots,n;
	\end{cases}
\end{align}
and,  by restriction to the paired diagonals,
	\begin{align}
	\omega_{k,n-k}(p_{\set n})=
	{\rm Alt}_1\cdots{\rm Alt}_{n-1}(-1)^{n-1}
	\sum_{\sigma\in\operatorname{cycl}(n)}
	\prod_{i=1}^n\widehat{\bK}(p_i,p_{\sigma(i)}),\qquad n\geq2,\label{eq:mixed-determinant}
\end{align}
where
\begin{align}\label{eq:mixed-kernel-cases}
	\widehat{\bK}(p_i,p_{\sigma(i)}) = \begin{cases}
		{\bK}(p_i,p_{\sigma(i)}) & i, \sigma(i) =1,\dots,k, \ i\not=\sigma(i); \\
		\bG_{x,y}^{(p_i)} {\bK}(p_i,p_{\sigma(i)}) & i=k+1,\dots,n,\ \sigma(i) =1,\dots,k; \\
		\bG_{x,-y}^{(p_{\sigma(i)})} {\bK}(p_i,p_{\sigma(i)}) & i=1,\dots,k,\ \sigma(i) =k+1,\dots,n; \\
		\bG_{x,y}^{(p_i)}\bG_{x,-y}^{(p_{\sigma(i)})} {\bK}(p_i,p_{\sigma(i)}) & i,\sigma(i) =k+1,\dots,n, \ i\not=\sigma(i).
	\end{cases}
\end{align}

\section{Hypergeometric BKP tau functions}\label{apx:HyperGeom}

In this appendix, we review topological recursion for the hypergeometric BKP tau function along the lines of~\cite{AS23}, which is related to a version of spin Hurwitz numbers alternative to the one considered in the present paper, in the framework of symplectic duality. As  a main application, we prove~\cite[Conjecture 6.1]{Alexandrov-BGW}.

The \emph{hypergeometric family of BKP tau functions}~\cite{Orlov} is defined by
\begin{equation}\label{eq:hypergeometric-BKP}
Z^r=\sum_{\lambda~\text{strict}}\Bigl(\prod\nolimits_{i=1}^{\ell(\lambda)}r_{\lambda_i}\Bigr)\frac{Q_\lambda(p)\;Q_\lambda(q)}{2^{\ell(\lambda)}}.
\end{equation}
We regard $p=(p_1,p_3,p_5,\dots)$ as the arguments of the tau function, while $r=(r_1,r_2,r_3,\dots)$ and $q=(q_1,q_3,q_5,\dots)$ are treated as parameters of the family (they can be specialized to arbitrary values such that $r_k\ne0$). This is indeed a BKP tau function for any parameter values satisfying $r_k\ne0$. Moreover, when $r_k=1$ for all $k$, we have
\begin{equation}\label{eq:Zr-at-r-equal-one}
Z^r\bigm|_{r_k=1}=\sum_{\lambda~\text{strict}}\frac{Q_\lambda(p)\;Q_\lambda(q)}{2^{\ell(\lambda)}}
=e^{2\sum_{k\in\Z^+_{\rm odd}}\frac{p_kq_k}k},
\end{equation}
which is manifestly a BKP tau function since its logarithm is bilinear in the two sets of times, and variation of the $r$-parameters is a BKP symmetry. The Taylor coefficients of $\log Z^r$ have a combinatorial interpretation as weighted spin Hurwitz numbers, alternative to~\eqref{eq:spinNumbers}, see e.g.~\cite{Lee,AS23,GKL}.

Let $\psi(\theta)$ be an even formal power series satisfying $\psi(0)=0$ and $\widetilde\psi(\theta)=\psi(\theta)+\sum_{k\ge1}\hbar^{2k}\psi_k(\theta)$ be its $\hbar$-deformation, with all $\psi_k(\theta)$ being even as well. We are interested in the following specialization of $Z^r$, see~\cite{AS23}:
\begin{equation}\label{eq:r-from-psi}
q_k=s_k/(2\hbar),\qquad
\log r_k=\frac{1}{2 \hbar}\int_{-k\hbar}^{k\hbar}\widetilde\psi(\theta)d\theta
=k\,\cS(2k\hbar\partial_{\theta})\,\widetilde\psi(\theta)\bigm|_{\theta=0},
\end{equation}
that is, we set
\begin{equation}\label{eq:OS-BKP}
Z^\psi=\sum_{\lambda~\text{strict}}e^{\frac1{2\hbar}\sum\limits_{i=1}^{\ell(\lambda)}
 \int\limits_{-\lambda_i\hbar}^{\lambda_i\hbar}\!\!\widetilde\psi(\theta)d\theta}
\;\frac{Q_\lambda(p)\;Q_\lambda(s/(2\hbar))}{2^{\ell(\lambda)}}.
\end{equation}

\begin{definition}\label{def:hypergeom-spectral-data}
\emph{The spectral curve data} associated with the series~\eqref{eq:OS-BKP} is the following collection of functions and differentials:
\begin{align} \label{eq:Xz}
y(z)&=\sum_{\substack{k\ge1\\k~\mathrm{odd}}}s_k\,z^k,
& X(z)&=z \, e^{-\psi(y(z))},
&  x(z)&=\log X=\log z-\psi(y(z)),
\end{align}
and
\begin{align}
 B(z_1,z_2)&=\frac{dz_1dz_2}{(z_1-z_2)^2}+\frac{dz_1dz_2}{(z_1+z_2)^2}.
\label{eq:B-hypergeom}
\end{align}
\end{definition}

Consider the $n$-point differentials $\omega^{(g)}_n$ associated with $Z=Z^\psi$ by~\eqref{eq:omegaZ}. One of the main results of~\cite{AS23} can then be stated as follows.

\begin{theorem}[A reformulation of~{\cite[Theorem 4.1; Propositions 4.3--4.5]{AS23}}]\label{thm:AS23-reformulation}
We have:
\begin{enumerate}
	\item  The differentials $\omega^{(g)}_n$ possess the topological expansion property; that is, they vanish unless $g$ is a nonnegative integer.
	\item Using~\eqref{eq:Xz} and~\eqref{eq:B-hypergeom}, the unstable differentials are given explicitly by
	\begin{equation}\label{eq:unstable-hypergeom}
		\omega^{(0)}_1(z)=y\,dx,\qquad \omega^{(0)}_2(z_1,z_2)=B(z_1,z_2).
	\end{equation}
	\item The symplectic duals of the system of differentials~$\omega^{(g)}_n$ with respect to the symplectic duality data $(x,y,\widetilde\psi)$ are trivial:
	\begin{equation}\label{eq:sympl-dual-trivial}
		\omega^{\dagger,(g)}_n=\delta_{(g,n),(0,1)}y_1\,\frac{dz_1}{z_1}+\delta_{(g,n),(0,2)} B.
	\end{equation}
\end{enumerate}
\end{theorem}
This interpretation differs substantially from the original formulation from~\cite{AS23}, since \emph{op.~cit.} adapted to the BKP context the results of~\cite{BDKS01,BDKS02} that were revisited by Hock and later on by the authors in~\cite{Hock-quantum-diLog,ABDKS05} in the realm of logarithmic topological recursion; the latter framework is applied here for reformulation.

The last assertion of Theorem~\ref{thm:AS23-reformulation} implies the following explicit closed formula for these differentials. Note that here we, again, rather use the framework of \cite{Hock-quantum-diLog,ABDKS05} adapted to the BKP context than the original approach in~\cite{AS23}; the sum over graphs that assembles commutation rules for the vertex operators is replaced by the determinantal formulas in the neutral free fermion representation:
\begin{align}
\bK^\dagger(z_1,z_2)&=\frac{z_1+z_2}{2\,(z_1-z_2)}\sqrt{\frac{dz_1}{z_1}}\sqrt{\frac{dz_2}{z_2}},\label{eq:bK-dagger}
\\
\bW^\dagger_n(z_{\set n},u_{\set n})&=\prod_{i=1}^n e^{u_i(\cS(u_i\hbar z_i\partial_{z_i})-1)y_i}
\\ \notag & \quad
 \times  {\rm Alt}_1\dots{\rm Alt}_{n-1}(-1)^{n-1}\sum_{\sigma\in {\rm cycl}(n)}\prod_{i=1}^n
\bK^\dagger(e^{\frac{u_i\hbar}{2}}z_i,e^{-\frac{u_{\sigma(i)}\hbar}{2}}z_{\sigma(i)}),\label{eq:bW-dagger}
\\\omega_n(z_{\set n})&=\prod_{i=1}^n\left(\sum_{r,j\ge0}
\bigl(d_i\tfrac{1}{dx_i}\bigr)^j\coeff{v}{j}
\biggl(\frac{\partial_{y}^re^{v\cS(v\hbar\partial_y)\widetilde\psi(y)}}
{e^{v\psi(y)}}\bigg|_{y=y_i}\biggr)
\coeff{u_i}{r}\right)\bW^\dagger_n
\\ \notag & \quad
+\frac{\delta_{n,1}}{\hbar}\sum_{j\ge1}
\bigl(d_1\tfrac{1}{dx_1}\bigr)^{j-1}\coeff{v}{j}
\biggl(\frac{e^{v\cS(v\hbar\partial_y)\widetilde\psi(y)}}
{e^{v\psi(y)}}\bigg|_{y=y_1}\biggr)dy_1, \label{eq:omega-from-dual}
\end{align}
where  $({\rm Alt}_i f)(z_i,u_i)=f(z_i,u_i)-f(-z_i,u_i)$.
The dual extended differentials $\bW^\dagger_n$ are computed here using the determinantal formulas of BKP integrability for the dual spectral curve with $x^\dagger=x+\psi(y)=\log z$.

The theorem above is stated in the setting of formal power series, with no convergence assumptions. Meanwhile, bypassing the reasoning given in~\cite{AS23}, the general properties of symplectic duality~\cite{ABDKS05} imply that
\begin{itemize}
	\item  under suitable meromorphy assumptions on $y(z)$ and $\psi(\theta)$,
	\item  with an appropriate choice of the $\hbar$-deformation $\widetilde\psi$ of~$\psi$,
	\item and suitable genericity conditions,
\end{itemize}
  the differentials $\omega^{(g)}_n$ satisfy topological recursion relations. The precise conditions on $y(z)$ and $\psi(\theta)$ may vary with the setting; see~\cite[Section 6]{ABDKS05} for details. Here is one possible setting (the so-called Family~I in the terminology of \emph{op.~cit.}).

\begin{theorem}\label{thm:family-I}
Assume that $y(z)$ is rational, that $\psi$ has the form $\psi(\theta)=P(\theta)+\log Q(\theta)$, where $P$ and $Q$ are even rational functions, and that
\begin{equation}\label{eq:psi-tilde-family-I}
\widetilde\psi(\theta)=P(\theta)+\frac{1}{\cS(\hbar\partial_\theta)}\log Q(\theta).
\end{equation}
If the coefficients of the rational functions $y(z)$, $P(\theta)$, and $Q(\theta)$ are in general position, then the differentials $\omega^{(g)}_n$ satisfy the relations of nondegenerate topological recursion on the rational spectral curve with affine coordinate $z$, with $(x,y)=(x(z),y(z))$ given by~\eqref{eq:Xz} and with $\omega^{(0)}_2$ given by~\eqref{eq:B-hypergeom}.
\end{theorem}

The case $Q=1$ is treated in~\cite{AS23} addressing a conjecture in~\cite{GKL}. The general theory of symplectic duality automatically extends the statement to arbitrary~$Q$. In particular, in the case $P=0$ this resolves~\cite[Conjecture 6.1]{Alexandrov-BGW}. To match precisely the formulation in \emph{op.~cit.} one has to notice that the spectral curve proposed there (as a classical limit of the quantum curve) canonically coincides with the one given in~\eqref{eq:Xz}.

Though the proof of this conjecture was at hand for quite some time, so far only a very special case was treated in the literature. See~\cite{WYZ-special-case} that addressed a very special case of \cite[Conjecture 6.1]{Alexandrov-BGW} that was earlier proposed in~\cite{Alexandrov-CandJ}.

Note that if the genericity condition for the rational functions $y(z)$, $P(\theta)$, and $Q(\theta)$ is not satisfied, then one should apply a suitable generalization of the original CEO topological recursion, cf.~Section~\ref{sec: various TR}. As we mentioned above, the most universal definition suitable for this application is the so-called \emph{Symplectic Topological Recursion (SymplTR)} that will be presented in a forthcoming paper~\cite{ABDKS-SymplTR}.


\printbibliography

\end{document}